# De l'impossibilité de l'expérience cruciale en physique : synthèse historique soutenant la critique par Duhem des conséquences de la mesure de la vitesse de la lumière dans l'eau.

**Olivier Morizot**
Aix-Marseille Univ, CNRS, Centre Gilles Gaston Granger, Aix-en-Provence, France
Olivier.morizot@univ-amu.fr

**Résumé**
En 1850, Foucault compare expérimentalement la vitesse de la lumière dans l'air à sa vitesse dans l'eau. Comme il est alors reconnu que la théorie ondulatoire et celle des projectiles encore en compétition prédisaient des résultats inverses à une telle mesure, Foucault « déclare le système de l'émission incompatible avec la réalité des faits ». Dans sa *Théorie Physique*, Duhem s'empare pourtant de cet exemple pour illustrer sa démonstration épistémologique de l'impossibilité de l'« experimentum crucis » en physique. Et l'ambition du présent article est d'augmenter la démonstration de Duhem de preuves historiques relatives à la question de la vitesse de la lumière dans l'eau. On passera effectivement en revue les quatre opinions majeures quant à la nature de la lumière envisagées de 1637 à 1801, menant chacune des théories concluant que le rapport des sinus est constant et égal au rapport des vitesses dans les deux milieux ; mais défendue chacune par au moins deux auteurs concluant à des rapports de vitesses inverses – donc à des prédictions inverses quant au résultat d'une éventuelle expérience de Foucault. La convergence des indices historiques confirmera ainsi que la nature de la lumière, considérée isolément, n'est en rien contrainte par la mesure de sa vitesse dans l'eau.

**Mots-clés**
Optique, réfraction, histoire, expérience cruciale, Foucault, Duhem

**Title**
Historical synthesis supporting Duhem's demonstration of the impossibility of the 'experimentum crucis' of comparison of the speed of light in air and water

**Abstract**
In 1850, Foucault was the first to compare experimentally the velocity of light in air to its velocity in water. And as the wave and projectile theories competing at the time predicted inverse results to such a measurement, Foucault "declared the emission system incompatible with the reality of the facts". Yet, in his *Physical Theory*, Duhem uses this very example to illustrate his epistemological demonstration of the impossibility of the 'experimentum crucis' in physics. Then, the ambition of the present article is to augment Duhem's demonstration with historical evidence relating precisely to the question of the speed of light in water. We will review the four major opinions on the nature of light developed between 1637 and 1801; all leading to theories concluding that the ratio of sines is constant and equal to the ratio of the velocities in the two media, but all being defended by at least two authors concluding to inverse ratios of those velocities – and thus to opposite predictions on the result of an experiment such as Foucault's. The convergence of historical evidence will then confirm that the nature of light, considered in isolation, is in no way constrained by the measurement of its speed in water.

Dans une note adressée le 3 décembre 1838 à l'Académie des Sciences, François Arago annonçait triomphalement :

> Je me propose de montrer dans cette Note comment il est possible de décider, sans équivoque, si la lumière se compose de petites particules émanant des corps rayonnans, ainsi que le voulait Newton, ainsi que l'ont admis la plupart des géomètres modernes ; ou bien si elle est simplement le résultat des ondulations d'un milieu très rare et très élastique, que les physiciens sont convenus d'appeler *éther*. Le système d'expériences que je vais décrire ne permettra plus, ce me semble, d'hésiter entre les deux théories rivales. Il tranchera *mathématiquement* (j'emploie à dessein cette expression), il tranchera mathématiquement une des questions les plus grandes et les plus débattues de la philosophie naturelle[1].

L'expérience imaginée alors par Arago, consistant à comparer la vitesse de la lumière dans l'air et dans l'eau à l'aide d'un dispositif à miroir tournant, ne sera pourtant réalisée qu'en avril 1850 par Léon Foucault[2]. Et au-delà de la prouesse technique et du résultat tant attendu que cette expérience offrait à la théorie physique[3], elle fut – avant même sa mise en œuvre – envisagée comme « cruciale ». Au sens où il était acquis qu'elle devait permettre de trancher définitivement entre théories « des projectiles » lumineux et théories « des ondulations » lumineuses ; et donc d'établir la nature véritable de la lumière. En effet, deux systèmes rivaux prétendaient encore expliquer alors les phénomènes lumineux[4]. Mais « Parmi ces phénomènes, l'un des plus simples et des plus apparents, la réfraction, résulte de deux actions opposées de la part des corps, suivant qu'on cherche à l'interpréter dans l'une ou dans l'autre théorie. D'après le système de l'émission, le changement de direction de la lumière serait dû à une accélération subie à son entrée dans les milieux réfringents. Dans le système des ondulations, ce même changement de direction devrait coïncider avec un ralentissement dans la vitesse de propagation du principe lumineux ». Ainsi est-il prédit que « l'un des deux succombera le jour où l'on constatera, par une expérience directe, dans quel sens se modifie la vitesse, lorsque la lumière pénètre d'un milieu rare dans un milieu plus dense, lorsqu'elle passe de l'air dans l'eau ou dans tout autre liquide ». Et de fait, l'expérience de Foucault montre que : « toujours la lumière se trouve retardée dans son passage à travers le milieu le plus réfringent. La conclusion dernière de ce travail consiste donc à déclarer le système de l'émission *incompatible* avec la réalité des faits[5] ».

Certainement est-il intéressant de souligner dès maintenant la prudence des conclusions de Foucault, relativement à la grandiloquence d'Arago : quand ce dernier annonce son ambition de déterminer « sans équivoque » la nature de la lumière ; l'autre conclue seulement avoir démontré l'incompatibilité d'un *système* avec « la réalité des faits ». Or les nuances subtiles entre ces deux positions seront justement au cœur de la démonstration que fera Pierre Duhem, un demi-siècle plus tard, de l'impossibilité de l'« *experimentum crucis* » en physique – s'appuyant justement sur l'exemple de cette expérience de Foucault. Les arguments épistémologiques avancés par Duhem[6] pour cette démonstration nous semblent incontestables, nous y reviendrons. Pour autant, puisque nous croyons que philosophie et histoire des sciences ne peuvent aller l'une sans l'autre, jugeons-nous intéressant d'étayer sa conclusion par une série d'éléments historiques supplémentaires qui viendront encore l'illustrer et la soutenir. Car en effet, un passage en revue de quatre options majeures envisagées pour décrire la propagation de la lumière entre 1637 et 1801 (par pression, par les voies les plus aisées, vibratoire et dynamique), mettant dos à dos, pour chacune de ces options, un couple d'auteurs majeurs parvenant à des conclusions opposées sur la question de la vitesse de la lumière dans l'eau, montrera en retour que la nature de la lumière n'est contrainte en rien par la mesure de cette vitesse ; et permettra donc de mieux saisir encore les raisons pour lesquelles une telle expérience ne peut être cruciale.

---

[1] ARAGO, « Système d'Expériences », 1838, p. 954.
[2] FOUCAULT, *Sur les vitesses relatives de la lumière dans l'air et dans l'eau,* 1853.
[3] Cette expérience est en effet réalisée presque deux siècles après la première estimation observationnelle de la vitesse de la lumière dans l'espace par RØMER, « Démonstration touchant le mouvement de la lumière », 1676.
[4] CANTOR, *Optics after Newton*, 1983. Les noms et le périmètre précis de ces deux catégories de théories, que nous emploierons tout au long du texte, sont d'ailleurs empruntés à ce même auteur.
[5] FOUCAULT, *Ibid.*, p. 34-35.
[6] DUHEM, *La théorie physique*, 1914, p. 285-289.

Annonçons toutefois avant d'aller plus avant que le problème et les éléments historiques convoqués ici sont bien connus des historiens de l'optique, et que seule la synthèse qui en est proposée a prétention à une quelconque originalité. Nous sommes néanmoins convaincus que ces éléments, en tant que tels, seront éclairants et utiles aux historiens d'autres champs de la science, aux étudiants de philosophie et plus encore aux scientifiques de profession qui gagneront incontestablement à se tourner vers nos revues pour alimenter une réflexion salutaire sur le fond historique et philosophique de leurs disciplines. C'est à l'attention de ces lecteurs en particulier que nous avons pris le risque de nous attarder sur l'explicitation des cas historiques évoqués, et de certains implicites propres à la communauté.

Mais gageons que pour tous, la synthèse de ces éléments d'histoire articulés à la question philosophique de l'impossibilité de l'expérience cruciale en physique aura un intérêt qui dépassera celui d'un simple catalogue. Que d'un point de vue historique, la mise en perspective de ces théories révèlera que c'est en général moins d'une hypothèse sur la nature de la lumière que d'une hypothèse mécanique relative au rapport de la lumière à la matière que dépend la conclusion sur le changement de vitesse de la lumière dans son passage de l'air à l'eau. Et que d'un point de vue philosophique, cette mise en contraste ravivera autant une réflexion sur l'importance de la pluralité en sciences, que sur le lien qu'entretient toute vérité scientifique aux hypothèses sur lesquelles elle repose, et sur les limites du processus de validation et d'invalidation de ces hypothèses par une expérimentation portant sur ses conséquences logiques.

# 1 Les théories de la pression lumineuse de Descartes et Hobbes

## 1.1 René Descartes

Le premier savant de la galerie que nous proposons de visiter se doit d'être René Descartes. A la fois parce qu'il est le premier à rendre publique une formulation correcte complète de la loi de la réfraction[7]. Mais aussi parce que ce faisant, il inscrit le problème de la réfraction dans le cadre d'une philosophie mécaniste supposant une nature matérielle de la lumière ou de son support, à laquelle souscriront l'ensemble des auteurs de la période considérée : qu'ils soient partisans des projectiles lumineux ou des vibrations d'un éther composé de minuscules particules, tous envisagent par défaut un support corpusculaire à la lumière, et postulent l'existence d'une mécanique régissant le mouvement de ces corpuscules et devant rendre compte des phénomènes lumineux. Quant au problème de la réfraction, les auteurs que nous évoquerons par la suite reconnaitront donc tous la même loi de la réfraction, mais se différencieront pourtant par leur manière de la justifier, toujours dans un cadre mécaniste[8].

---

[7] De fait, de nombreux auteurs se sont penchés sur la réfraction avant Descartes (Ptolémée, Ibn al-Haytham, Vitelion, Pecham, Kepler…). Certains parvenant même avant lui à une formulation équivalente de la loi de la réfraction (Ibn Sahl, Harriot, Snell). Mais c'est bien à la formulation de Descartes que l'ensemble des auteurs ultérieurs vont se référer et c'est pourquoi nous commencerons avec lui.

[8] La conception mécaniste de la lumière est cependant incompatible avec quantité de théories de la lumière et de la vision précédant celles de Descartes ; que l'on pense aux théories antiques du rayon visuel (SIMON, *Le regard, l'être et l'apparence*, 1988), à la nature quasi-spirituelle des rayons lumineux d'Ibn al-Haytham ou de Kepler (SIMON, *Archéologie de la vision*, 2003), pour réaliser l'homogénéité mécaniste des théories de la lumière depuis Descartes jusqu'au milieu du XIXᵉ siècle, et pour justifier l'intérêt de se limiter ici à cet ensemble, suffisamment

Enfin, Descartes sera un cas exemplaire relativement à notre propos, du fait que la conclusion à laquelle il parvient quant à la vitesse de la lumière dans l'eau est souvent présentée comme une conséquence nécessaire de son mode démonstration de la loi des Sinus, de la nature qu'il attribue à la lumière, ou des deux – entretenant dès lors l'idée qu'une mesure de cette vitesse aurait pu invalider le modèle. L'illusion étant entretenue par la structure surprenante de la justification mécanique qu'il offre à sa loi de la réfraction à l'endroit même où il la présente. Certains indices tendent cependant à montrer que chez Descartes, comme chez la grande majorité des auteurs que nous évoquerons, l'opinion quant à la vitesse de lumière dans l'eau est bien antérieure à toute démonstration sur la réfraction ; et que celle-ci fait office d'hypothèse rectrice pour la construction du modèle proposé expliquer ce phénomène – ou d'élément de confirmation *a posteriori* de la plausibilité du modèle – plutôt que de conclusion nécessaire de celui-ci.

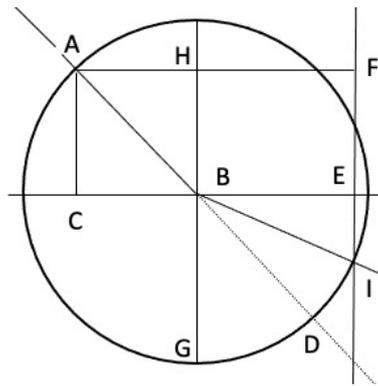

**Fig. 1 :** DESCARTES, *Dioptrique*, 1902, p. 98. Une hypothétique balle arrive selon la direction (*AB*) à l'interface entre l'air et l'eau. La direction de la balle est alors modifiée ; et au lieu de se propager jusqu'en *D*, comme si elle poursuivait son mouvement dans l'air, elle est déviée vers *I*

En quelques mots, la justification cartésienne de la réfraction pourrait être présentée ainsi : la sensation de lumière est la conséquence d'une pression exercée sur la rétine. Celle-ci résulte d'une sorte de force centrifuge produite par le tourbillonnement incessant des corpuscules de matière ignée composant les corps lumineux (premier élément), exercée sur les corpuscules de matière subtile (second élément) qui sont à leur contact, puis propagée de proche en proche au sein de cette matière qui remplit complètement les cieux et vient même se glisser dans les moindres pores des corps transparents[9]. La théorie cartésienne de la lumière est donc résolument une théorie du « milieu », où la propagation de la lumière se fait sans véritable mouvement, mais par transmission d'une pression, ou d'une « inclination à se mouvoir », à travers un milieu matériel. Inclination à se mouvoir seulement, puisque l'espace cartésien étant par définition plein et compact, les corpuscules d'éther ne peuvent se déplacer et se contentent de presser les uns sur les autres ; c'est même ce qui permet de justifier d'une transmission instantanée, selon Descartes, de la lumière. Toutefois, c'est à l'aide d'une analogie avec le mouvement d'une balle, donc selon un modèle généralement associé aux théories de l'« émission » de corpuscules lumineux, qu'il justifie pour la première fois la loi de la réfraction dans sa *Dioptrique* (Fig. 1) :

> Pensons maintenant que la balle, qui vient d'*A* vers *D*, rencontre au point *B* […] de l'eau, dont la superficie *CBE* lui ôte justement la moitié de sa vitesse, […] je dis que cette balle doit passer de *B* en ligne droite, non vers *D*, mais vers *I*.

---

cohérent pour permettre de relever le rôle fondamental de certains détails. On notera d'ailleurs que Kepler, envisageant un moment la possibilité d'une loi de la réfraction régie par un rapport de sinus constant, évacue rapidement cette option, au titre qu'elle nécessiterait d'envisager la lumière comme étant faite de projectiles matériels – ce qui pour lui était insoutenable (KEPLER, *Paralipomènes à Vitelion*, 1980, pp. 210-211).
[9] DESCARTES, *Le Monde*, 1909, p. 84-97.

> *[…] Or […], puisque la balle qui vient d'A en ligne droite jusques à B, se détourne étant au point B, et prend son cours de là vers I, cela signifie que la force ou facilité, dont elle entre dans le corps CBEI, est à celle dont elle sort du corps ACBE, comme la distance qui est entre AC et HB, à celle qui est entre HB et FI, c'est-à-dire comme la ligne CB est à BE.*
>
> Enfin, d'autant que l'action de la lumière suit en ceci les mêmes lois que le mouvement de cette balle, il faut dire que, lorsque ses rayons passent obliquement d'un corps transparent dans un autre, qui les reçoit plus ou moins facilement que le premier, ils s'y détournent en telle sorte, qu'ils se trouvent toujours moins inclinés sur la superficie de ces corps, du côté où est celui qui les reçoit le plus aisément, que du côté où est l'autre : et ce, justement à proportion de ce qu'il les reçoit plus aisément que ne fait l'autre[10].

En d'autres termes, la loi de la réfraction ici formulée pour la première fois, peut être symbolisée par l'égalité des rapports $\frac{CB}{BE} = \frac{f_2}{f_1}$, où $f_1$ et $f_2$ représentent cette « force » ou cette « facilité » avec laquelle la balle se propage dans chacun des deux milieux. Ou encore $\frac{\sin i}{\sin r} = \frac{f_2}{f_1}$, en notant $i$ l'angle entre la direction d'incidence de la balle et la droite orthogonale au point d'impact, $r$ l'angle entre la direction de la balle réfractée et cette même droite, et en appelant Sinus d'un angle la longueur de la demi-corde portée par celui-ci[11]. Ce à quoi il faut nécessairement ajouter que pour parvenir à cette formulation, Descartes a préalablement affirmé que la vitesse de la balle dans un milieu, ou la facilité de la lumière à le traverser, ne dépend pas de sa direction de propagation ; donc que le rapport $\frac{f_2}{f_1}$, auquel est égal le rapport des Sinus, est constant.

Descartes introduit et justifie donc la réfraction de la lumière par analogie avec le mouvement hypothétique d'une balle projetée obliquement à travers l'air vers la surface de l'eau et telle « que ni la pesanteur ou légèreté de cette balle, ni sa grosseur, ni sa figure, ni aucune autre telle cause étrangère ne change son cours[12] ». Selon ce modèle, au passage de l'air à l'eau, la vitesse de la balle décroît d'un facteur $\frac{v_{eau}}{v_{air}}$ indépendant de l'obliquité de son arrivée ; mais la composante de la vitesse qui est parallèle à la surface n'est pas affectée par le changement de milieu, car elle ne rencontre aucune résistance dans cette direction à l'instant de la réfraction ; soit en formalisme moderne : $v_{air} . \sin i = v_{eau} . \sin r$. Et la combinaison de ces deux hypothèses mène naturellement à la conclusion : $\frac{\sin i}{\sin r} = \frac{v_{eau}}{v_{air}} = constante$.

Ce faisant, Descartes semble forcé d'associer deux hypothèses qui sembleront contradictoires à nombre de ses lecteurs : Descartes se voit en effet obligé d'affirmer, afin de justifier ce choix d'un modèle de projectiles pour justifier de la déviation de la pression lumineuse, (1) que l'« inclination à se mouvoir, que j'ai dit devoir être prise pour la lumière, doit suivre en ceci les mêmes lois que le mouvement[13] » – à ceci près seulement que l'inclinaison à se mouvoir se propage instantanément. Mais constatant bien que, selon son modèle, la balle déviée par l'eau doit s'écarter de la normale – alors que la lumière pénétrant dans l'eau s'en rapproche – Descartes doit immédiatement concéder une exception à cette règle et défendre (2) que, contrairement au mouvement lui-même, l'inclinaison à se mouvoir se propagera d'autant plus aisément que le milieu sera dense[14].

---

[10] DESCARTES, *Dioptrique*, 1902, p. 98-101. Le passage en italique, correspondant à la formulation de la loi de la réfraction, a été souligné par nous. Une explication pédagogique de la démonstration de la loi de la réfraction par Descartes est développée par SABRA, *Ibid.*, 1981, p. 99-116.
[11] Le « sinus » généralement utilisé aujourd'hui n'étant que ce même Sinus évalué dans un cercle de rayon unité.
[12] DESCARTES, *Ibid.*, 1902, p. 99.
[13] DESCARTES, *Ibid.*, 1902, p. 89. L'argument selon lequel l'inclinaison à se mouvoir suit exactement la même voie selon laquelle cette même action déplacerait le premier de ces corps si les autres n'étaient pas sur son chemin étant développé par ailleurs (DESCARTES, *Le Monde*, 1909, p. 102-103).
[14] DESCARTES, *Dioptrique*, 1902, p. 102-103.

Ainsi, a-t-on a souvent suggéré que l'affirmation par Descartes d'une plus grande « facilité » des milieux denses à laisser passer la lumière était la conséquence nécessaire du modèle mécanique qu'il avait choisi pour justifier la nouvelle loi de la réfraction[15] : ayant découvert que le rapport des Sinus est égal à une constante ; disposant d'un modèle mécanique permettant de justifier l'égalité du rapport des Sinus dans le cas du mouvement d'une balle impondérable mais prédisant l'augmentation de l'angle de réfraction dans le cas où elle perdrait « la moitié de sa vitesse » en pénétrant dans le second milieu ; Descartes semble ici contraint de déduire *in extremis* que la lumière – qui s'approche de la normale lorsqu'elle passe de l'air à l'eau – se propage plus « aisément » dans l'eau que dans l'air. Ou plus « vite », si l'on s'autorise à associer à ce concept de « force ou facilité » avec laquelle la lumière passe dans les milieux un concept de « vitesse » : $\frac{\sin i}{\sin r} = \frac{f_2}{f_1} = \frac{v_{eau}}{v_{air}}$, et puisque $i$ est supérieur à $r$, $v_{eau}$ doit être supérieur à $v_{air}$[16]. Et cette impression a certainement contribué à entretenir l'opinion que la conclusion sur le rapport des vitesse de la lumière dans l'air et dans l'eau était une conséquence nécessaire de leurs modèles, à laquelle les théoriciens successifs avaient été contraints de se plier afin d'en assurer la cohérence ; donc que la comparaison expérimentale de ces vitesses pourrait rigoureusement permettre de les invalider logiquement.

Pourtant, comme le souligne A. I. Sabra[17], on trouve dans l'un des carnets de notes remplis par Descartes sur la période allant approximativement de 1619 à 1621 l'entrée suivante :

> Puisque la lumière ne peut être générée que dans la matière, elle est plus facilement générée là où il y a plus de matière, toutes choses étant égales par ailleurs ; elle pénètre donc plus facilement dans un milieu plus dense que dans un moins dense. D'où vient que la réfraction se fait dans celui-ci à l'opposé de la perpendiculaire, dans l'autre vers la perpendiculaire[18].

Si l'on tient alors pour acquis que Descartes ne disposait pas encore de la loi de la réfraction – et donc encore moins de sa justification – à l'époque où il écrivait ces mots[19], on doit accepter que la condition pour les vitesses à laquelle il souscrit n'est pas la conséquence inéluctable de son modèle de la réfraction, mais bien plutôt un postulat qui le précède et qu'il estime compatible avec une simple observation du phénomène. Surtout, il devient clair que son modèle pour la justification de la loi de la réfraction n'a pas imposé à Descartes sa conclusion sur cet étonnant rapport des vitesses. Et l'on peut imaginer au contraire que son postulat premier sur les vitesses aura justement participé à formuler son modèle, ou tout du moins à en crédibiliser des conclusions étonnantes pour certains, mais qui confirmaient pour lui le postulat d'origine.

---

[15] SABRA, *Theories of light,* 1981, p. 106-107.
[16] De fait, Descartes n'emploie le mot « vitesse » qu'à trois reprises dans les pages dédiées à l'explication de la réfraction – toujours quand il s'agit de parler de la fameuse balle dont le mouvement est d'abord pris pour modèle (DESCARTES, *Ibid.*, 1902, p. 97-98). Mais au moment de passer à l'expression de la loi même, le terme disparait : un changement terminologique s'impose, puisque Descartes soutient l'idée que la vitesse de la lumière est infinie. Ainsi, s'il est déjà difficile dans ce cas de parler de « vitesse » de la lumière, devient-il impossible d'envisager rigoureusement son accélération ou son ralentissement au passage dans l'eau. D'où la précaution de langage prise par Descartes. Toutefois, Fermat et Huygens interprèteront explicitement cette proposition de Descartes en termes de changement de vitesse ; et Fermat utilisera lui-même indifféremment les termes « force » et « vitesse » dans sa propre interprétation du phénomène. Pratique qui ne souleva pas d'objection de Descartes et fut même adoptée par ses partisans (SABRA, *Theories of light,* 1981, p.113). Aussi, puisqu'il ne s'agit pas ici de percer le message cartésien mais de l'articuler aux autres théories de son siècle, nous autorisons-nous ici – comme cela a été fait à l'époque – à associer ce changement de « force ou facilité » à traverser un milieu – qui justifie pour Descartes la réfraction une notion de vitesse de la lumière – à un changement de « vitesse », quitte à attribuer à ce mot un sens plus vague que sa définition cinématique contemporaine.
[17] SABRA, *Ibid.,* 1981, p. 105-106.
[18] DESCARTES, *Cogitationes Privatae*, 1908, p. 242-243.
[19] Gaston Milhaud estime que cette démonstration de la loi de la réfraction est postérieure à 1628 (MILHAUD, *Descartes Savant*, 1921, p. 109).

Ce qui nous parait intéressant dans ce premier exemple, c'est donc que la condition sur le rapport des vitesses de la lumière dans l'air et dans l'eau semble relever d'un postulat, plutôt que de la conclusion rigoureusement déduite de la nature que l'auteur attribue à la lumière. Étant entendu que ce postulat n'est pas purement arbitraire et qu'il doit évidemment être plausible relativement à cette nature, il ne revêt toutefois pas ce caractère d'absolue nécessité associé à la conclusion d'une déduction logique, qui fait que son invalidation peut remettre en cause l'hypothèse dont elle découle. A ce titre, une mesure expérimentale invalidant la condition des vitesses défendue par Descartes ne saurait invalider son pari sur la nature de la lumière. Entendons-nous bien : il va de soi que la mesure même d'une vitesse finie de la lumière, aurait remis en cause l'hypothèse de l'instantanéité de sa progression et aurait possiblement mené Descartes « à confesser que toute [sa] Philosophie était entièrement renversée[20] ». Toutefois, il semble que l'évolution des formules employées par Descartes pour décrire cette instantanéité au fil de ses écrits laisse une place à l'éventualité d'une lumière se propageant simplement trop vite pour que l'on puisse le constater[21]. Et s'il n'a pu être témoin de la démonstration de la vitesse finie de la lumière par Rømer[22], sa théorie optique continua bien d'être défendue par ses partisans. Certains refusèrent un temps ce résultat – signe, s'il en fallait, du caractère non univoque des résultats expérimentaux. Mais, plus significativement, d'autres concédèrent que la dioptrique cartésienne pouvait s'en accommoder sans changement majeur, à condition par exemple de renoncer à l'incompressibilité rigoureuse des corpuscules d'éther ou à la plénitude de l'espace[23]. Dès lors peut-on imaginer qu'une mesure de la vitesse de la lumière dans l'eau réalisée à la même époque aurait de la même manière imposé aux cartésiens d'abandonner le choix initial du rapport des vitesses fait par Descartes – et aurait donc rigoureusement invalidé la théorie cartésienne originale –, mais les aurait toutefois autorisé à la réécrire en préservant la nature de pression transmise dans un milieu continu attribuée à la lumière, en ajustant simplement la démonstration de loi de la réfraction en conséquence[24].

Enfin s'il serait impossible – et sans grand intérêt d'ailleurs – de démontrer que Descartes lui-même aurait su s'accommoder d'une autre condition des vitesses, nous montrerons maintenant qu'un autre auteur majeur de la période, plaidant pour une même nature de la lumière, a produit une justification cohérente et semblablement plausible de la loi des Sinus, postulant toutefois un rapport des vitesses inverse de celui défendu par Descartes. Ce faisant, nous confirmerons qu'aucune conclusion sur la vitesse de la lumière dans l'eau ne découle nécessairement de l'hypothèse selon laquelle la lumière serait une pression instantanément propagée dans un milieu plein.

### 1.2 Thomas Hobbes

En effet, sur la période s'étalant de 1630 à 1651, Thomas Hobbes développe une théorie optique mécaniste reposant sur un modèle de la lumière très explicitement comparable à celui proposé par Descartes[25]. Mais si Hobbes reconnait la validité de la loi de la réfraction cartésienne dès qu'il en prend connaissance – et s'il défend comme Descartes l'idée d'un espace matériel plein et incompressible, dans lequel la lumière se propage instantanément par communication d'une pression exercée par les corps lumineux sur les corpuscules d'éther voisins, puis entre eux, de proche en proche, jusqu'à la rétine – il rejette résolument l'option prise par Descartes de justifier la réfraction (et la réflexion) par analogie entre la propagation de cette pression et celle d'un corps en mouvement : puisque ce ne sont manifestement pas les mêmes choses, il faut les traiter de manière explicitement différenciée. Hobbes se propose donc d'élaborer une description cinématique complète de la propagation d'une impulsion dans un milieu continu et compact qui puisse justifier de la propagation rectiligne, de la réflexion et de la réfraction de la lumière. Et il y parvient manifestement ; à condition d'admettre que la lumière se propage plus aisément dans l'air que dans l'eau.

---

[20] Comme il le suggère lui-même dans cette lettre à Beeckmann où il tente d'ailleurs de le convaincre de l'inutilité d'une telle expérience, au titre du fait que les observations astronomiques des éclipses de Lune auraient déjà incontestablement démontré l'instantanéité de la propagation lumineuse (DESCARTES, « A Monsieur *****, Lettre XVII, Version », 1659, p. 140).
[21] RODIS LEWIS, « Quelques remarques », 1998.
[22] RØMER, « Démonstration touchant le mouvement de la lumière », 1676.
[23] Comme par exemple HUYGENS, *Traité de la lumière*, 1690, ou LE CAT, *Traité des sens*, 1744, p. 91-92.
[24] D'autant que l'irruption de multiples hypothèses *ad hoc* au fil de cette démonstration semble offrir au modèle une souplesse suffisante pour pouvoir l'adapter à la condition des vitesses inverse sans avoir à renoncer aux fondements de la théorie.
[25] SHAPIRO, « Kinematic Optics », 1973 ; BERNHARDT, « Hobbes et le mouvement de la lumière », 1977.

Selon la première proposition du *Tractatus opticus*[26], un corps lumineux exerce une poussée sur les corpuscules de matière subtile qui l'entourent, en se dilatant par un mouvement et simultané du centre vers toute sa périphérie, puis se rétracte alternativement, selon un processus régulier (mais pas forcément périodique) et d'amplitude si faible qu'elle est insensible. Et à l'instant même où la source commence à se dilater, l'œil subit aussitôt un début de pression, quel que soit son éloignement : sous l'effet de la dilatation, la première couche périphérique du milieu incompressible prend instantanément la place de la seconde, et ainsi de toutes les suivantes jusqu'à la rétine. Comme chez Descartes, la sensation de lumière résulte donc d'une pression qui se transmet instantanément dans le milieu, sans transport de matière.

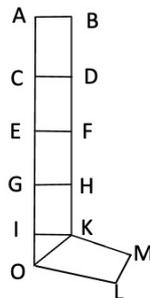

**Fig. 2 :** HOBBES, *Tractatus opticus,* 1644, p. 570. La ligne de lumière *AB* se propage perpendiculairement aux droites *AI* et *BK* qui marquent sa progression. Le rayon lumineux droit est le volume limité par le rectangle *ABIK*. Sa réfraction à l'interface *KO* entraine la rotation de la ligne de lumière, en conservant la perpendicularité et sans modification de la largeur du rayon, contrairement ce que la Figure, plus naïve que celles qu'il emploie par la suite, devrait suggérer. La ligne de lumière se propage alors à une vitesse constante, mais différente, dans le second milieu. Cette vitesse dépendant de la densité de celui-ci

Cependant, pour sa justification mécanique du rapport constant des Sinus lors de la réfraction, Hobbes ne considère pas la déviation d'un corpuscule en mouvement, mais celle d'une « ligne de lumière » : c'est-à-dire une couche infime du milieu en mouvement, orthogonale à celui-ci, et limitée latéralement à une largeur « inférieure à quelque magnitude donnée que ce soit[27] ». Cette ligne de lumière envisagée par Hobbes est droite et se propage perpendiculairement à ce que *nous* appelons rayons lumineux. De fait, ces demi-droites que nous appelons rayons lumineux n'ont pas de réalité physique pour lui : ce ne sont que des lignes abstraites permettant de représenter le chemin parcouru par le front lumineux. L'entité physique véritable à considérer dans la théorie optique de Hobbes est donc soit la ligne de lumière elle-même, soit la portion d'espace balayée par cette ligne depuis son point de départ – et que lui appelle « rayon »[28]. Dès lors (Fig. 2), « Un rayon droit est un rayon qui, si on le coupe par un

---

[26] HOBBES, *Tractatus opticus*, 1644, p. 568. Noter que le traité de Hobbes est rédigé vers 1640, donc après la publication de la *Dioptrique.*
[27] HOBBES, *Ibid.*, 1644, p. 575.
[28] « Puisque dans les faits, un rayon est un chemin selon lequel un mouvement est projeté depuis un corps lumineux, il ne peut qu'être le mouvement d'un corps ; il suit qu'un rayon est le lieu d'un corps, et par conséquent qu'il possède trois dimensions. Un rayon est donc un espace solide. » (HOBBES, *Ibid.*, 1644, p. 571)

plan passant par son axe, est un parallélogramme (comme *AK*). Un rayon réfracté est un rayon formé de deux rayons droits faisant un angle, avec une partie intermédiaire[29] ». Et la réfraction de ce rayon se voit expliquée par une différence de résistance à la progression de la ligne de lumière dans deux milieux de densités différentes : « J'appelle « plus rare » un milieu qui résiste moins à la réception du mouvement, et « plus dense » un qui est plus résistant. De plus, Je suppose que l'air est plus rare que l'eau, l'eau plus rare que le verre, et le verre plus rare que le cristal[30] ». Ce en quoi Hobbes s'oppose donc diamétralement à Descartes, et qui annonce déjà un rapport des vitesses inverse pour justifier du rapport constant des Sinus.

La démonstration la loi de la réfraction proposée par Hobbes est alors la suivante[31] : le rayon lumineux épais et solide considéré est par essence limité à l'avant par la ligne de lumière qui doit toujours être orthogonale à la poussée – c'est-à-dire à la direction de propagation de la lumière – et sa largeur doit toujours être conservée[32]. Or, si ce rayon arrive obliquement à la surface séparant les deux milieux, l'un des côtés du rayon (soit l'une des extrémités de la ligne de lumière) pénètre dans le second milieu avant l'autre, et sa poussée s'y voit frustrée d'autant que le second milieu est plus dense que le premier. La ligne de lumière, supposée rigide, pivote alors, jusqu'à ce que son autre extrémité pénètre dans le second milieu. Et le rayon se courbe en conséquence se rapprochant de la normale, comme on s'y attend lors de son passage de l'air à l'eau. La loi de la réfraction avancée par Descartes se trouve ainsi démontrée, à condition de supposer une résistance plus grande de l'eau ($R_{eau}$) à la progression de la lumière : $\frac{\sin i}{\sin r} = \frac{R_{eau}}{R_{air}}$. Rapport qui est affirmé constant en vertu du fait que le rapport de ces résistances est supposé l'être[33]. Et que, selon le même raisonnement que nous avons appliqué plus tôt à Descartes, on peut[34] reformuler comme un rapport de vitesses : $\frac{\sin i}{\sin r} = \frac{v_{air}}{v_{eau}}$ – dont on remarque qu'il est direct cette fois, et donc inverse à celui proposé par Descartes.

Si les convictions mécanistes de Hobbes ne doivent rien à Descartes, son idée d'une lumière se propageant instantanément sous l'effet d'une pression communiquée à un milieu compact lui est certainement empruntée[35], de même que l'est évidemment la loi des Sinus elle-même. Toutefois, malgré la similitude manifeste de ces deux manières d'envisager la nature de la lumière, et malgré le choix commun de justifier cette même loi par une différence de résistance du milieu à la progression de la lumière – que l'on a pu réinterpréter en termes de différence de vitesse de celle-ci dans chacun des milieux – on voit que l'un et l'autre arrivent à des conclusions sur la résistance de l'eau à la progression de la lumière qui sont cohérentes avec leurs systèmes respectifs, et pourtant diamétralement opposées. La correspondance qui suivra entre ces deux auteurs par l'intermédiaire de Mersenne s'attardera d'ailleurs significativement sur cette opposition, au cours de laquelle Hobbes n'aura de cesse de critiquer les prétendues preuves empiriques d'une résistance diminuant avec la dureté du milieu traversé que Descartes tente d'apporter[36].

---

[29] HOBBES, *Ibid.*, 1644, p. 571.
[30] HOBBES, *Ibid.*, 1644, p. 567, hypothèse 5.
[31] On trouvera une explication de cette démonstration dans SHAPIRO, « Kinematic Optics », 1973, p. 258-263.
[32] La conservation de la largeur du rayon n'est pas postulée, mais manifestement appliquée dans le raisonnement.
[33] HOBBES, *Tractatus opticus,* 1644, p. 581.
[34] Avec toutes les précautions qu'il se doit pour évoquer ce système à la propagation de la lumière est supposée instantanée, comme le fait par exemple SHAPIRO, « Kinematic Optics », 1973, p. 153-156.
[35] BERNHARDT, « Hobbes et le mouvement de la lumière », 1977, p. 16.
[36] La difficulté principale dans la résolution de ce débat, comme le souligne Jean Bernhardt (« Hobbes et le mouvement de la lumière », 1977, p. 18), provenant probablement « du caractère délibérément fictif ou de la rationalité subjective des explications avancées par Descartes dans la *Dioptrique*, où il n'est question de rendre compte des « propriétés » de la lumière qu'« en la façon qui me semble la plus commode », sans égard à sa « nature ». »

Quoi qu'il en soit, nous sommes bien là face à deux conceptions presque identiques de la nature de la lumière, et à deux démonstrations du même rapport constant des Sinus d'incidence et de réfraction, parvenant à deux conclusions inverses quant à la facilité pour la lumière à traverser l'air et l'eau. Constat qui – dès lors que l'on s'autorise le lien entre facilité de traversée d'un milieu et vitesse – nous permet d'entretenir l'opinion qu'une théorie de la lumière de type cartésienne aurait pu s'accommoder indifféremment d'un rapport des vitesses, ou de son inverse, pour justifier du rapport constant des Sinus. Que dès lors, l'ambition de produire une expérience cruciale permettant de trancher quant à l'hypothèse spécifique de la nature de la lumière par la comparaison de ses vitesses dans l'air et dans l'eau était effectivement aussi illusoire que l'a démontré Duhem dans sa *Théorie Physique*[37]. Quand bien même le modèle localement développé par Descartes pour expliquer les ressorts de la réfraction aurait-il été invalidé par l'expérience, la théorie de Hobbes – identique sur tant de points que le premier n'a eu de cesse de reprocher à l'autre de l'avoir plagié[38] – ne l'aurait pas été. Suggérant bien que l'expérience n'aurait été en mesure ni de contester la nature de la lumière défendue par ces deux auteurs ; ni même d'empêcher une théorie cartésienne alternative de se reconfigurer autour de cette donnée.

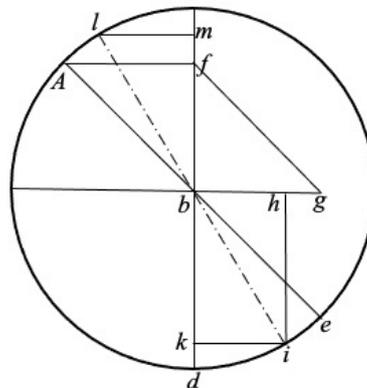

**Fig. 3 :** HOBBES, *De Corpore*, 1655, p. 215. Représentation cinématique de la réfraction d'une inclination au mouvement issue de *A* et rencontrant la surface *bg* séparant l'air de l'eau en *b*, pour être ensuite réfractée vers *i*. Le rayon du cercle marque la distance parcourue par l'inclination en un temps donné

On objectera peut-être à cela que la démonstration de la réfraction ici proposée par Hobbes est très éloignée de celle proposée par Descartes. Mais quelques années plus tard seulement, Hobbes développera une nouvelle démonstration de la réfraction, formellement semblable à celle de Descartes, mais reposant toujours sur un rapport des vitesses inverse[39]. Sans qu'il soit aisé de déterminer la raison exacte qui a motivé le développement de cette démonstration alternative[40], Hobbes se propose d'étudier la réfraction d'un « mouvement » ou d'une « inclination » à se mouvoir, et pose comme préalable deux hypothèses diamétralement opposées à celles adoptées par Descartes, à savoir : (1) que c'est la composante parallèle à la surface de l'inclination au mouvement qui sera modifiée par la réfraction, (2) que cette inclination à se mouvoir se propagera plus lentement dans un milieu plus dense. Le rapport des vitesses dans les deux milieux $\frac{v_{air}}{v_{eau}}$ n'est pas explicite dans la démonstration, mais on sait qu'il est égal au rapport inverse des densités des milieux $\frac{\rho_{eau}}{\rho_{air}}$, lui-même égal au rapport des distances parcourues parallèlement à la surface dans le premier et le second milieu (à savoir l'air puis l'eau) dans un même temps donné. Ainsi (Fig. 3) : $\frac{v_{air}}{v_{eau}} = \frac{\rho_{eau}}{\rho_{air}} = \frac{Af}{bh} = \frac{\sin i}{\sin r}$. Ce modèle cinématique justifie donc toujours le rapport des Sinus avant et après réfraction par un rapport de vitesses ; mais un rapport inverse à celui suggéré par Descartes.

---
[37] DUHEM, *La théorie physique*, 1914, p. 285-289.
[38] SHAPIRO, « Kinematic Optics », 1973, p. 160.
[39] HOBBES, *De Corpore*, 1655, p. 217-219.
[40] Voir l'enquête que Hao Dong a menée sur cette question (DONG, « Hobbes's model of refraction », 2021).

S'il est vrai que ce modèle règle le problème des deux hypothèses apparemment peu compatibles de la justification cartésienne il en introduit toutefois d'autres, non moins surprenantes[41]. Mais ce que l'existence même de ce modèle concurrent – et pourtant si proche formellement – de celui de Descartes démontre à nos yeux, c'est que ni la nature de la lumière, ni même le formalisme choisi pour décrire sa propagation, n'imposaient véritablement dans ce cas de conclusion nécessaire quant au rapport des vitesses de la lumière dans l'air et dans l'eau. Ce dont on peut déduire en retour qu'une expérience de mesure de vitesse de la lumière aurait certes pu invalider la théorie optique cartésienne sous la forme qu'on lui connaît. Mais que, même si rien ne permet de prédire la réaction de Descartes à l'éventualité d'une mesure de la vitesse dans l'eau réalisée de son vivant – et si à vrai dire on s'en moque – l'exemple offert ici par Hobbes prouve qu'une théorie cartésienne alternative *aurait pu* se recomposer, en abandonnant l'hypothèse posée *a priori* d'une plus grande facilité de la lumière à se propager dans l'eau, et en s'accommodant de corrections mineures qui compenseraient cet abandon. Et qu'en particulier, ni la nature attribuée à la lumière, ni même le formalisme utilisé pour décrire sa réfraction n'auraient nécessairement eu à être abandonnés. A condition bien sûr qu'on ait continué de leur attribuer une plausibilité supérieure à celle des systèmes concurrents.

Mais la force d'une thèse en histoire des sciences se mesurant au nombre des indices convergents venant l'étayer, nous tâcherons dans les pages qui suivent de la renforcer selon un même schéma : passant successivement en revue les principales autres options envisagées à l'époque quant à la nature de la lumière – et les principaux formalismes physico-mathématiques employés pour justifier de sa réfraction –, puis confrontant dans chaque cas deux auteurs ayant conclu inversement quant au rapport des vitesses de la lumière dans l'air et dans l'eau.

## 2 Les théories de la voie la plus aisée de Fermat et Leibniz

### 2.1 Pierre de Fermat

Dans une lettre de 1664, Pierre de Fermat décrit les raisons qui très tôt l'avaient mené à rejeter la démonstration cartésienne de la réfraction[42]. Premièrement, elle ne reposait selon lui que sur des comparaisons ; or en géométrie plus encore qu'ailleurs, les comparaisons ne font pas des preuves. Deuxièmement, elle présupposait que le passage de la lumière était plus aisé dans les milieux les plus denses, « ce qui semble choquer le sens commun ». Enfin, elle supposait injustement que le mouvement horizontal de la lumière n'était pas modifié lors de la réfraction[43]. Or, même si Fermat a pu romancer ce récit *a posteriori*[44], il est avéré qu'après sa première lecture de la *Dioptrique* et pendant les vingt ans qui suivirent – et même après qu'elle ait été confirmée expérimentalement – celui-ci refusa de reconnaître la validité de la loi des Sinus. Et, convaincu qu'une démonstration fausse n'avait pu que mener à une loi erronée – même si elle pouvait suffisamment s'approcher de la vérité pour que l'expérience ne l'en puisse discerner – il finit par se lancer à son tour à la recherche de la véritable loi de la réfraction. Mais en se basant pour sa part sur un principe si fondamental et vrai et sur une méthode si rigoureuse que la vérité ne pourrait lui échapper. Le principe en question est celui selon lequel la nature agit toujours selon les voies « les plus aisées »[45]. Et la méthode est celle qu'il a développée quelques vingt ans plus tôt pour le calcul des minima et maxima.

---

[41] Notamment le fait qu'une analyse de la Figure 3 en termes de vitesse suggère que la vitesse totale est globalement inchangée (les distances *Ab* et *bi* étant traversées dans un même temps), ou que la vitesse perpendiculaire à la surface doit augmenter alors que la vitesse parallèle diminue (SHAPIRO, « Kinematic Optics », 1973, p. 171).
[42] FERMAT, « Lettre de Fermat à M. de ****, 1664 », 1894.
[43] On a vu que cette même proposition fut également remise en cause par Hobbes dans son *De Corpore*.
[44] SABRA, *Theories of light*, 1981, p. 117-126.

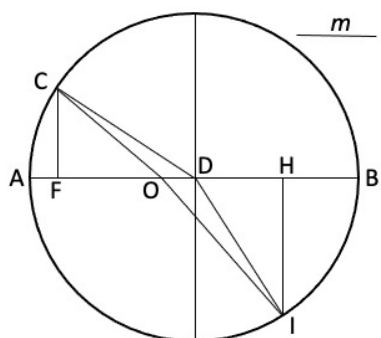

**Fig. 4 :** FERMAT, « Analyse pour les réfractions », 1896. Le problème consiste à évaluer le chemin le plus court pour aller de *C* à *I* en considérant les vitesses de la lumière au-dessus et en-dessous de l'interface *AB* dans un rapport connu

Mis ensemble, ce principe et cette méthode permettent à Fermat de déterminer dans un premier temps le chemin « le plus aisé » pour qu'un rayon lumineux passant par le point *C* (Fig. 4) – situé dans un milieu « rare » où la résistance à sa propagation est faible – rejoigne un point *I* quelconque – situé dans un milieu « dense » où la résistance à sa propagation est plus forte[46]. La procédure revient alors à déterminer la position du point *D*, situé à l'interface entre les milieux, par lequel doit passer la lumière pour que la résistance totale opposée à sa propagation sur le chemin allant de *C* à *I* soit minimale. Et pour Fermat, cela revient formellement à minimiser la quantité $R_1.CD + R_2.DI$, où $R_1$ et $R_2$ quantifient la « résistance » des deux milieux à la propagation de la lumière. Processus qui lui permet de conclure mathématiquement que ce chemin ainsi défini comme « le plus aisé » – que la lumière se doit *par principe* d'emprunter – respecte la loi des sinus : $\frac{\sin i}{\sin r} = \frac{R_2}{R_1}$. Mais cette première démonstration ne statue pas explicitement sur la vitesse de la lumière, même si, en affirmant que la résistance $R_2$ de l'eau à la progression de la lumière est supérieure à celle $R_1$ de l'air, elle suggère en effet que la lumière se propage plus lentement dans l'eau que dans l'air.

Quelque temps plus tard, Fermat s'attèle à la résolution du problème inverse, consistant à montrer que le chemin déterminé par la loi des sinus est le plus court en temps pour joindre un point *M* à un point *N* situé dans un milieu où la résistance au mouvement est plus grande[47]. Ce deuxième essai commence par les mots suivants :

> Le savant Descartes a proposé pour les réfractions une loi qui est, comme on dit, conforme à l'expérience ; mais, pour la démontrer, il a dû s'appuyer sur un postulat absolument indispensable à ses raisonnements, à savoir que le mouvement de la lumière se ferait plus facilement et plus vite dans les milieux denses que dans les rares ; or ce postulat semble contraire à la lumière naturelle.
> En cherchant, pour établir la véritable loi des réfractions, à partir du principe contraire, – à savoir que le mouvement de la lumière se fait plus facilement et plus vite dans les milieux rares que dans les denses, – nous sommes retombés précisément sur la loi que Descartes a énoncée[48].

Ainsi, Fermat met-il lui-même en question l'idée que l'option envisagée par Descartes pour la nature de la lumière – ou le modèle qu'il avait choisi pour expliquer la réfraction – l'aurait contraint dans sa détermination du rapport des facilités de la lumière à traverser l'air et l'eau ou, comme Fermat le reformule, des vitesses de propagation de la lumière dans les deux milieux. Selon le récit proposé ici, Descartes aurait d'abord postulé que la lumière se propageait « plus vite » dans l'eau que dans l'air ; et de là se serait affairé à démontrer sa loi de la réfraction. De même que c'est en partant du principe contraire que Fermat affirme être parvenu lui-même à la démontrer. D'après lui, l'idée du rapport des vitesses de la lumière est donc un postulat qui précède dans ces deux cas le modèle d'explication de la réfraction.

---

[45] PARMENTIER, « Le principe de simplicité des voies », 1995 ; RASHED, « Fermat et le principe du moindre temps », 2019.
[46] FERMAT, « *Analyse pour les réfractions* », 1896.
[47] Pour une explication complète de cette démonstration en deux temps de la loi de la réfraction, voir JIMENEZ C. A. & al., « On the analytic and synthetic demonstrations in Fermat's work on the law of refraction », 2019.
[48] FERMAT, « *Synthèse pour les réfractions* », 1896, p. 151-152.

Dans ce cas, une invalidation expérimentale du postulat sur le rapport des vitesses de la lumière n'invaliderait ni le formalisme choisi pour décrire la réfraction – qui lui est postérieur ; ni le postulat posé sur la nature de la lumière – dont il n'est pas déduit logiquement, et avec lequel il suffit qu'il puisse plausiblement coexister. Concrètement, chez Descartes, la plus grande vitesse de la lumière dans l'eau n'est effectivement pas démontrée logiquement à partir de la nature supposée de la lumière ; celle-ci est au contraire supposée *a priori*, et considérée comme d'autant plus plausible qu'elle est compatible avec la nature supposée de la lumière[49]. Mais chez Hobbes, la même nature supposée de la lumière soutenait la plausibilité de l'option inverse pour les vitesses, prouvant s'il le fallait que plausibilité n'est pas nécessité.

Et dans le cas de la démonstration de la réfraction par le principe des « voies les plus aisées » le même schéma semble à l'œuvre : si dans l'esprit de Fermat ce principe vient soutenir le postulat selon lequel la lumière se propage plus rapidement dans l'air que dans l'eau, nous allons voir que pour Leibniz il est en mesure de s'accorder avec le postulat inverse.

### 2.2 Gottfried Wilhelm Leibniz

Dans son *Unicum opticae, catoptricae et dioptricae principum,* publié en 1682, Gottfried Leibniz, emploie non seulement un formalisme mathématique très proche de celui Fermat, reposant sur un calcul de minima et de maxima, pour démontrer la loi de la réfraction[50]. Mais ce « principe unique », qu'il entend appliquer à « l'optique, la catoptrique et la dioptrique », est également un principe de la voie « la plus aisée » : « La lumière émise d'un foyer parviendra au point à éclairer par le chemin le plus facile de tous ». Principe interprété toutefois comme un principe de moindre *résistance* – plutôt que de moindre *temps* – et selon lequel : « Les difficultés du chemin dans les différents milieux sont en raison composées et de la longueur des chemins et de la résistance des milieux ». Incidemment, Leibniz parvient donc à une même formulation du problème que Fermat dans son *Analyse pour les réfractions* : le chemin emprunté par la lumière pour aller de *C* à *I* est celui passant par le point *D* situé sur le dioptre et tel que $R_1.CD + R_2.DI$ doit être minimal, où $R_i$ quantifie toujours la résistance du milieu au passage de la lumière. Et la résolution de ce problème mène évidemment à la loi des sinus : $\frac{\sin i}{\sin r} = \frac{R_2}{R_1}$. Mais si Leibniz suppose, comme Fermat, une résistance plus grande dans l'eau que dans l'air, il lui semble plus raisonnable d'en conclure que la lumière est plus rapide dans l'eau : la lumière, constituée de projectiles lumineux, se propagerait plus rapidement dans les milieux les plus denses, dont les particules matérielles plus rapprochées empêcheraient la diffusion latérale des corpuscules lumineux, ce qui accélèrerait leur flux[51].

Qu'une telle conclusion lui ait été permise met en évidence le fait que l'interprétation par Fermat du principe des voies « les plus aisées » en principe de « moindre temps » n'était également qu'un choix particulier ; que la résolution du problème par calcul du chemin opposant le moins de « résistance » à la lumière – tel que Fermat l'avait lui-même formulé initialement[52] – menait à la loi correcte des sinus, sans pour autant impliquer un ralentissement de la lumière dans l'eau – dès lors que l'on admettait comme Descartes et Leibniz que l'eau offre certes une plus grande résistance à la lumière, mais qu'une plus grande résistance implique une accélération de celle-ci.

---

[49] DESCARTES, *Cogitationes Privatae*, 1908, p. 242-243.
[50] LEIBNIZ, *Unicum opticae, catoptricae et dioptricae principum*, 1958. Voir également McDONOUGH, « Leibniz and optics », 2018.
[51] LEIBNIZ, *Ibid.*, 1958, p. 189-190. Une analogie pouvant être établie entre la progression de la lumière d'un milieu moins résistant à un milieu plus résistant et l'accélération d'un flux d'eau passant d'un tuyau à large section à un tuyau plus étroit. Cette analogie présentant d'ailleurs le mérite, par rapport à celle des projectiles, de proposer une explication mécanique plausible à l'éventuelle accélération de la lumière au moment du changement de milieu.

Nous venons donc de voir que, dans le cas des quatre théories présentées jusqu'ici, la condition sur les vitesses de la lumière n'est pas déduite de raisonnements purement mathématiques ou optiques, mais seulement posée et acceptée en vertu de sa plausibilité à l'égard de l'hypothèse retenue pour décrire la nature de la lumière. Mais peut-être a-t-on également remarqué que dans chacune de ces théories – même dans celles de Leibniz et Fermat, qui sont probablement les moins mécaniques de celles que l'on étudiera – le problème de la réfraction était formulé *a priori* en des termes mécaniques, introduisant les concepts – relativement vagues – de « dureté », de « densité » ou de « résistance » pour qualifier la propriété supposée des corps matériels rendant compte de leur action sur la lumière qui les traverse. Ce qu'apporte ce nouveau constat, c'est que la plausibilité attribuée *a priori* à la condition des vitesses de la lumière, dérive seulement d'une analogie mécanique : la vitesse de la lumière dans les différents milieux est posée en vertu de l'analogie établie par chaque auteur avec le comportement supposé de l'entité mécanique associée à la lumière (projectiles lancés, propagation d'une pression dans un milieu plein, flux d'un fluide…) lors de la traversée de milieux matériels plus ou moins « denses », « durs » ou « résistants ». C'est ce qui fait qu'ici non plus, ni le principe des voies les plus aisées, ni le formalisme physico-mathématique de calcul des minima et maxima, n'imposaient de conclusion nécessaire quant à la vitesse de la lumière dans l'eau. En d'autres termes, une comparaison des vitesses de la lumière dans l'air et dans l'eau eût-elle été réalisée à l'époque, qu'elle n'aurait pas été en mesure d'écarter ni la possibilité que la lumière soit une pression se propageant dans un milieu compact, ni qu'elle soit une entité matérielle quelconque empruntant les voies les plus aisées – ceci, quoi que l'expérience eût donné.

Leibniz abandonnera pourtant sa théorie de la lumière quelques années plus tard seulement, en faveur de celle développée par Christiaan Huygens[53]. Celui-ci ne proposait pourtant pas d'expérience nouvelle ; mais il avait développé un modèle vibratoire de la lumière qui était en mesure de rendre compte mécaniquement non seulement des phénomènes les plus couramment observés[54], mais aussi du phénomène relativement nouveau de double réfraction. Incidemment, Huygens démontrait aussi que le principe des voies les plus aisées pouvait être considéré comme une conséquence de son modèle vibratoire, à condition d'adopter la condition des vitesses défavorable à Leibniz. Nous analyserons donc dans la prochaine partie ce qui a pu suggérer une supériorité de la théorie hugonienne de la réfraction sur celles qui l'ont précédée. Nous le ferons cependant après avoir rapidement envisagé la théorie vibratoire de Robert Hooke, qui l'avait précédée d'un quart de siècle.

---

[52] On note d'ailleurs que dans l'*Analyse pour les réfractions* de Fermat, le mot « temps » n'apparait pas. Ce n'est que dans sa *Synthèse pour les réfractions*, rédigée ultérieurement, que Fermat s'emploie explicitement à démontrer, en partant de la loi des sinus et en admettant que la lumière – modélisée comme un « mobile » – est plus rapide dans l'air que dans l'eau, que le chemin emprunté par la lumière lors des réfractions est le plus court en temps.

[53] LEIBNIZ, « Lettre de Leibniz à Huygens du 26 avril 1694 », 1901.

[54] A savoir la propagation en ligne droite et non instantanée de la lumière, la possibilité pour les rayons lumineux de se croiser sans s'empêcher, la réflexion et la réfraction. Précisons toutefois que Huygens n'emploie jamais le terme « vibration », mais celui de « percussions » sans « suite réglée » pour décrire le mouvement des corpuscules lumineux, ou d'« onde » pour décrire la mise en branle de l'éther qui en découle (HUYGENS, *Traité de la lumière*, 1690, p. 4). Comme annoncé en introduction, je parlerai néanmoins dans son cas – comme dans celui de Hooke – de « vibration » et de « théorie vibratoire », conformément à la classification précise des théories de la lumière proposée par CANTOR, *Optics after Newton*, 1983. Classification qui, si elle court le risque d'être rétrospective, offre toutefois un moyen efficace de comparaison des théories optiques contemporaines et postérieures à Newton.

## 3 Les théories vibratoires de Huygens et Hooke

### 3.1 Robert Hooke

L'exemple de Robert Hooke sera en effet intéressant à analyser en premier lieu, pour ce qu'il articule des hypothèses empruntées à Descartes et Hobbes[55] en un nouveau modèle vibratoire préfigurant celui de Huygens[56], mais soutenant, contrairement à ce-dernier, l'idée d'une accélération de la lumière dans l'eau.

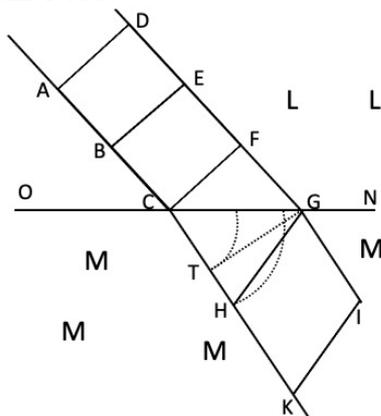

**Fig. 5** : HOOKE, *Micrographia*, 1665, Planche VI. Le front d'onde initialement orthogonal aux rayons (*CF*) s'incliner par rapport à ceux-ci (*GH*) du fait de l'entrée de la lumière en *C* dans un milieu où elle se propage plus rapidement, alors qu'elle continue à se propager lentement de *F* à *G*

Hooke considérait que la lumière était causée par un mouvement vibratoire très rapide, de faible amplitude et régulier des différentes parties des corps lumineux[57]. Cette vibration étant communiquée aux corpuscules du milieu avoisinants et de ceux-ci, via leurs voisins, de proche en proche jusqu'à la rétine. La propagation de ce « mouvement vibratoire » n'est pas nécessairement instantanée[58], mais se fait à vitesse égale dans toutes les directions quand le milieu est homogène ; résultant en la production d'une onde sphérique, longitudinale car produite par les chocs des corpuscules dans la direction de propagation de la vibration, centrée sur le point émetteur, et dont la surface coupe perpendiculairement ce que l'on avait jusque-là appelé « rayons lumineux »[59]. Toutefois, ceci ne vaut que dans les milieux homogènes ; car d'après Hooke, la spécificité de la réfraction – qui fait qu'elle est à l'origine de la production des couleurs que l'on voit après les prismes notamment – réside justement dans le fait qu'elle produit une inclinaison du front d'onde par rapport aux rayons ; le front d'onde restant toutefois cette surface jusqu'à laquelle la vibration a été propagée en un temps donné. En effet, selon Hooke, la vibration pénétrant dans l'eau au point *C* va s'y propager plus « aisément » qu'elle ne continuera de le faire entre *F* et *G* (Fig. 5). Ce qui résultera en la production d'un front d'onde *GH* incliné par rapport aux rayons ; inclinaison à laquelle l'œil sera sensible et qui sera à l'origine des couleurs produites lors de certaines réfractions. Le côté *CHK* de l'impulsion lumineuse, qui est le plus avancé donc, produit la sensation de bleu ; le côté *GI*, selon lequel la lumière est en retard, produit la sensation de rouge ; les autres couleurs n'étant que des dérivés ou des mélanges de ces deux couleurs, qui sont les pôles fondamentaux de la théorie hookienne de la couleur[60].

---

[55] SHAPIRO, « Kinematic Optics », 1973, p. 202-207.
[56] HUYGENS, « Projet du Contenu de la Dioptrique », 1916, p. 742, note 1 ; SHAPIRO, *Ibid.*, 1973, p. 205 ; SABRA, *Theories of light*, 1981, p. 186-187.
[57] C'est manifestement une amélioration apportée par Hooke aux théories du milieu, dont la plus avancée était alors celle de Hobbes, envisageant le corps lumineux comme vibrant dans sa globalité, de la même manière que bat le cœur. Et cette amélioration se retrouvera chez Huygens. Par ailleurs, notons que Hooke ne s'exprime pas sur la périodicité de ces vibrations, que Huygens écartera pour sa part explicitement.
[58] Hooke remarque par exemple que l'argument en faveur de la propagation instantanée avancé par Descartes, reposant sur l'observation des éclipses de Lune, est faussé puisqu'il présuppose ce qu'il est censé prouver, et n'écarte pas rigoureusement l'éventualité que la lumière soit plus rapide que ce que l'on peut observer par cette méthode. Huygens en développera la démonstration dans son *Traité de la lumière*, 1690, p. 4-6. Leurs deux théories démontrent dès lors la possibilité relativement peu coûteuse d'envisager une vitesse finie de propagation de la lumière dans un espace plein.
[59] HOOKE, *Micrographia*, 1665, p. 54-57.

Quant au cadre strict de notre article toutefois, on remarquera que Hooke ne propose pas à proprement parler de démonstration de la loi de la réfraction : ce n'est pas le problème qu'il s'agit de traiter dans son texte. Il se contente d'y admettre la loi de la réfraction telle que formulée par Descartes, pour s'affairer ensuite à expliciter l'effet produit par la réfraction sur l'onde qui se propage, qui selon lui fournit la clé de compréhension des couleurs prismatiques. Ainsi, de même qu'il affirme qu'« en accord avec le plus précis et excellent des Philosophes *Des Cartes,* je suppose que le sinus de l'angle d'inclinaison dans le premier *milieu* est au sinus de réfraction dans le second comme la densité du premier l'est à la densité du second[61] », Hooke admet avec le même Descartes que la vitesse de progression de la lumière est inversement proportionnelle à cette densité, et donc que la lumière voyage plus vite ou plus aisément dans l'eau que dans l'air : $\frac{\sin i}{\sin r} = \frac{d_{eau}}{d_{air}} = \frac{v_{air}}{v_{eau}}$.[62] Et c'est une fois ces deux éléments posés qu'il s'engagera dans une démonstration géométrique entièrement dédiée à déterminer l'inclinaison du front d'onde[63].

Bien qu'il ne s'agisse pas pour lui d'expliquer la raison de ce lien entre densité du milieu et vitesse de la lumière, Hooke ne manque pas d'en formuler une justification mécanique plausible :

> Ce n'est pas mon affaire en ce lieu que d'établir les raisons pour lesquelles tel ou tel corps devrait entraver plus les Rayons, et d'autres moins : comme pourquoi l'eau devrait transmettre les Rayons plus aisément, quoique plus faiblement que l'air. Tout ce que je suggèrerai en général à ce sujet sera que je suppose que le *milieu* MMM possède moins de la matière subtile ondulante, et que cette matière est moins impliquée par elle, alors que je suppose que LLL contient une plus grande quantité de la substance fluide ondulante, et qu'elle est plus impliquée avec les particules de ce *milieu*[64].

Cette position parait d'autant plus surprenante – et donc intéressante à analyser – qu'elle s'oppose à celle traditionnellement associée aux théories de la vibration[65]. D'autant que l'hypothèse inverse sur les vitesses aurait justement permis à Hooke de démontrer la loi des sinus (ce qu'il omet de faire), et de maintenir en même temps la perpendicularité du front d'onde avec les rayons qu'il avait explicitement affirmée dans sa première description des vibrations lumineuses. Mais comme souvent en histoire des sciences, ce qui semble un *hiatus* depuis notre époque ne relève ni d'une étourderie de l'auteur, ni d'une incohérence de son système, mais bien d'un choix délibéré et justifié de sa part : Hooke connaît la théorie de Hobbes qui justifie précisément la réfraction par un ralentissement du front de lumière pénétrant dans l'eau, combiné à la nécessité de maintenir la perpendicularité entre ce front et les rayons lumineux. Or, si Hooke rejette cette solution, c'est précisément parce que, ne produisant pas d'effet physique notable sur le front de l'onde, elle n'est pas à même de justifier cette « modification » de la lumière qui doit nécessairement expliquer l'apparition des couleurs du prisme :

> Cette *Hypothèse* que l'industrieux *Mersenne*[66] a publiée au sujet du mouvement plus lent de la terminaison du Rayon dans un *milieu* plus dense que dans un plus rare et fin semble tout compte fait insuffisante pour résoudre une abondance de *Phénomènes*, dont un qui n'est pas le moins considérable, qu'il est impossible par cette supposition que la moindre couleur soit générée par la réfraction des Rayons ; car puisque par cette *Hypothèse* l'*impulsion ondulante* est toujours transportée perpendiculairement, ou à angles droits avec le Rayon ou la Ligne de direction, il suit que l'impact de l'impulsion lumineuse, après qu'elle ait été une ou deux fois réfractée (à travers un Prisme, par exemple), doit affecter l'œil avec la même sorte d'impact que si elle n'avait pas été réfractée du tout[67].

---

[60] HOOKE, *Ibid.*, 1665, p. 64 ; BLAY, « Un exemple d'explication mécaniste au XVIIe siècle », 1981, p. 109.
[61] HOOKE, *Ibid.*, 1665, p. 57.
[62] Où $d_{eau}$ et $d_{air}$ sont des densités que l'on pourrait qualifier d'*optiques* – et non pas pesantes – des deux milieux, lesquelles quantifient justement la facilité des milieux transparents à être traversés.
[63] HOOKE, *Ibid.*, 1665, p. 57. On pourra se référer aussi à un commentaire de cette démonstration par Sabra (*Theories of light*, 1981, p. 192-194) ou Shapiro (« Kinematic Optics », 1973, p. 195-197). On y remarquera que la construction de la réfraction par Hooke est remarquablement similaire à celle proposée par Huygens : chaque partie du faisceau étant passée dans le second milieu s'y propageant effectivement à la vitesse de la lumière dans ce milieu indépendamment du fait que les parties toujours présentes dans le premier continuent de s'y propager à une autre vitesse (construction que l'on ne trouvait pas chez Hobbes qui considérait « la ligne de lumière » comme toujours rectiligne). Si bien que seul le choix du rapport des vitesses semble pouvoir les différencier.
[64] HOOKE, *Ibid.*, 1665, p. 57-58.
[65] Les plus connues en tout cas – comme celles de Huygens que nous allons voir ensuite, mais aussi d'Euler, Young ou Fresnel – affirment en effet au contraire une progression plus lente de l'onde lumineuse dans l'eau.

Hooke n'écarte donc pas l'hypothèse d'un ralentissement de la lumière dans l'eau par inattention, par fidélité aveugle à Descartes, ou par inconscience de ses implications pour une théorie vibratoire ; mais par choix, et en toute connaissance de cause. Ce choix n'est évidemment pas une conséquence nécessaire de sa démonstration de la réfraction puisqu'il n'en propose pas ; ni de la nature vibratoire qu'il attribue à la lumière puisque d'autres y associeront la condition inverse. Il est significatif de l'importance supérieure que Hooke attribuait à la résolution du problème des couleurs du prisme, plutôt qu'à celui de la réfraction qu'il pouvait alors légitimement considérer comme résolu – au moins du point de vue de la production d'une loi décrivant le phénomène – et qui à ce titre l'intéressait probablement moins. Mais ce choix n'est pas non plus une conséquence nécessaire de la solution que Hooke propose au problème des couleurs prismatiques : ce qu'il s'agit de mettre en évidence pour lui c'est une modification de la lumière incidente, qui selon lui réside dans une inclinaison du front d'onde par rapport aux rayons, laquelle est en effet naturellement justifiable par une différence de vitesse de propagation de l'onde dans les deux milieux. Mais puisque l'inclinaison calculée du front d'onde ne mène ensuite à aucune conséquence nécessaire observable, cette inclinaison aurait tout aussi bien pu être défendue pour un rapport $\frac{v_{eau}}{v_{air}}$ quelconque, y compris inférieur à 1, dès lors que ce rapport aurait été différent du rapport des sinus. Car c'est là que réside la seule nécessité pour Hooke : que la condition sur les vitesses aujourd'hui traditionnellement associée aux théories vibratoires ne soit pas respectée.

Et s'il est indéniable que la théorie cartésienne fournissait alors une expression simple de ce rapport des vitesses qui convenait à la démonstration d'une inclinaison, il est manifeste que, comme dans les théories déjà évoquées, le rapport des vitesses dans la théorie de Hooke relève moins d'une déduction logique rigoureuse que d'un choix délibéré entre deux options qu'il a minutieusement examinées ; puis parmi lesquelles il a manifestement choisi *a priori* celle qui lui paraissait la plus plausible au regard de la conception qu'il se faisait de la propagation de la lumière dans les milieux.

### 3.2 Christiaan Huygens

Huygens envisage la lumière comme le résultat de percussions irrégulières et rapides des corpuscules des corps lumineux, qui se communiquent à des corpuscules de matière subtile durs et élastiques répartis densément dans tout l'espace et susceptible de transmettre cette vibration sur de très grandes distances, à une vitesse considérable mais finie. Et la propagation de chaque vibration à vitesse uniforme dans toutes les directions se manifeste, selon lui, par la progression dans le milieu éthéré d'un front d'onde longitudinale sphérique et centré sur le corpuscule. Mais si l'on examinait le phénomène de plus près, on verrait que ce front d'onde résulte en réalité de la vibration à l'unisson des corpuscules d'éther situés à la surface tangente aux ondelettes sphériques produites par chacun des corpuscules d'éther précédemment mis en vibration par sa propre rencontre avec le front d'onde. C'est là ce que l'on appelle aujourd'hui le « principe de Huygens » : le front d'onde macroscopique, propagé depuis l'émetteur à vitesse finie, est aussi la tangente à l'ensemble des ondelettes créées à des temps antérieurs par la mise en vibration de chaque corpuscule du milieu, et qui se sont propagées à la même vitesse. Ainsi, comme Huygens le confesse dans une lettre à Colbert, l'idée même de « vitesse finie de la lumière » – à laquelle il consacre une très longue partie de son *Traité* – est la clé de voute de sa théorie :

> J'ai vu depuis peu avec bien de la joie la belle invention qu'a trouvée le Sr. Römer, pour démontrer que la lumière en se répandant emploie du temps […] cette démonstration m'a agréée d'autant plus que […] j'ai supposé la même chose touchant la lumière, et démontré par là les propriétés de la réfraction[68].

---

[66] Rappelons que le *Tractatus opticus* de Hobbes avait été publié à la fin d'une compilation de textes réalisée par Mersenne. La mention de la « terminaison du Rayon » faite ensuite par Hooke ne laisse pourtant pas de doute quant au fait qu'il s'agit de la théorie de Hobbes – articulée autour du concept de « front de lumière » – qui est évoquée ici.
[67] HOOKE, *Ibid.*, 1665, p. 100.

Premier savant à élaborer une théorie de la lumière postérieure à la démonstration de Rømer – et donc fermement armé d'un concept véritable de vitesse de propagation de la lumière – Huygens peut sereinement affirmer que cette vibration de l'éther qu'est la lumière se propage plus lentement dans les corps les plus denses. Il le justifie par un long argumentaire envisageant plusieurs options, mais favorisant celle selon laquelle le mouvement des corpuscules d'éther se transmettrait aux corpuscules de matière, dont le « ressort un peu moins prompt que n'est celui des particules éthérées » permettrait un « progrès des ondes de lumière […] plus lent au-dedans [du] corps, qu'[…] au dehors[69] ». Et cette affirmation, à nouveau posée *a priori* et justifiée seulement par une analogie mécanique, mène en effet à une démonstration complète de la réfraction ; démonstration qui – on l'a évoqué – permet également de retrouver le principe de moindre temps de Fermat[70] et peut donc être considérée comme plus fondamentale que lui.

On connaît bien cette démonstration[71] : considérant le cas simple d'un front d'onde incident plan *CA*[72] arrivant obliquement en faisant un angle *i* avec la surface *AB* séparant l'air de l'eau (Fig .6), et supposant que la lumière se propage moins vite dans le second milieu, Huygens explique que l'ondelette produite par l'arrivée du front d'onde sur le dioptre en *A* ne progressera que jusqu'à la surface sphérique *SNR* pendant le temps que prendra l'onde incidente pour terminer sa progression dans le premier milieu, de *C* à *B*. Fixant le rapport $\frac{v_{air}}{v_{verre}}$ à 3/2 – qui se trouve opportunément être le rapport des Sinus pour la réfraction de l'air au verre commun – Huygens peut même situer précisément l'ondelette *SNR* au moment où la lumière arrive finalement en *B*, et répéter le même protocole pour tous les points *K* intermédiaires. Et puisqu'il a été affirmé que le front d'onde global était toujours la surface tangente aux ondelettes intermédiaires, Huygens démontre que le front réfracté est le segment *BN*[73], incliné par rapport au dioptre d'un angle tel que $\frac{\sin ABN}{\sin BAC} = \frac{v_{verre}}{v_{air}}$. Mais comme les rayons lumineux « ne sont autre chose que les lignes droites suivant lesquelles les parties des ondes s'étendent[74] », ils sont conséquemment perpendiculaires au front de l'onde, et l'on trouve : $\frac{\sin i}{\sin r} = \frac{v_{air}}{v_{verre}}$.

---

[68] HUYGENS, « Lettre De Huygens à Colbert, 14 octobre 1677 », 1899. On peut même deviner de ce qui précède que le traitement rigoureux que fait Huygens de la vitesse finie de la lumière est équivalent à son fameux principe.
[69] HUYGENS, *Traité de la lumière*, 1690, p. 30.
[70] HUYGENS, *Ibid.*, 1690, p. 39-41.
[71] HUYGENS, *Ibid.*, 1690, p. 33-36. On pourra voir aussi : SABRA, *Theories of light*, 1981, p. 198-230, ou SHAPIRO, « Kinematic Optics », 1973, p. 207-258 pour un commentaire plus développé de cette démonstration.
[72] La planéité du front d'onde est rigoureusement justifiée par la section relativement faible de la surface d'onde examinée relativement à l'immense distance du point émetteur (HUYGENS, *Ibid.*, 1690, p. 33).
[73] Au lieu du front d'onde hypothétique *BG* si les ondes se propageaient à la même vitesse dans les deux milieux.

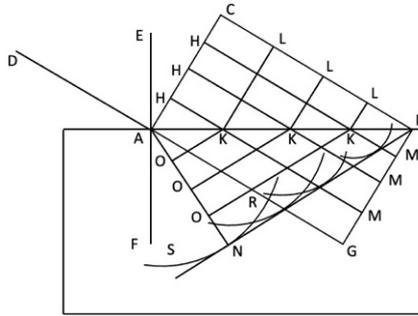

**Fig. 6 :** HUYGENS, Traité de la lumière, 1690, p. 33. Un front d'onde incident plan (*CA*), orthogonal aux rayons exciter une ondelette en *A* qui se propage plus lentement dans l'eau que ne continue de le faire la lumière progressant de *C* à *B* dans l'air. Au moment où le faisceau considéré commence à se propager tout entier dans l'eau, le nouveau front d'onde (*BN*), tangent à l'ensemble des ondelettes créées successivement depuis chaque point où l'onde atteignait le dioptre, est plan. Et les rayons lumineux réfractés – marquant la trajectoire de la lumière et toujours orthogonaux à ce front d'onde – respectent strictement la loi des sinus

La démonstration de Huygens a ceci d'admirable que – à condition d'admettre l'équivalence entre le front d'onde principal et la tangente commune aux ondelettes intermédiaires – elle propose en effet un modèle cinématique complet et cohérent justifiant la réfraction. Elle pose certes *a priori* l'hypothèse d'une vitesse de la lumière plus faible dans l'eau que dans l'air. Mais puisque cette hypothèse permet la démonstration rigoureuse de la loi, il est aussi bien possible d'envisager la condition sur les vitesses comme étant une prémisse *nécessaire* à la démonstration : si l'on retourne le sens de la démonstration pour remonter de l'observation expérimentale de la loi des Sinus à ses causes, la condition hugonienne des vitesses semble bien émerger comme une *conséquence* nécessaire de *cette* théorie vibratoire de la lumière. Celle-ci lie directement et univoquement la réfraction à la vitesse de la lumière dans les milieux, sans l'intervention d'hypothèse *ad hoc* intermédiaire[75].

Pour autant, Robert Hooke ne manquera pas de souligner que si la théorie de Huygens est capable de justifier la loi des Sinus, elle reste parfaitement incapable d'expliquer les couleurs prismatiques, et ne peut par conséquent être raisonnablement acceptée. Sans être nécessairement fausse, elle est « insuffisante […] à moins qu'il n'y ait quelque autre propriété omise de ces Rayons ou ondulations, qui doit être découverte et ajoutée et ce qui est déjà dit d'eux avant qu'elle soit admise[76] ». Hooke est d'ailleurs loin d'être le seul à ne pas accepter l'évidence trompeuse de la théorie de Huygens : nombreux sont ceux qui lui reprocheront son incapacité à donner une justification dynamique de l'accélération de la lumière ressortant de l'eau ; à esquisser la moindre explication des couleurs[77] ; ou ne serait-ce qu'à proposer un modèle acceptable de la propagation rectiligne de la lumière[78]. Ce que soulève ce témoignage

---

[74] HUYGENS, *Ibid.*, 1690, p. 35.
[75] A l'exception évidemment du principe des ondelettes, c'est-à-dire de l'équivalence entre l'onde macroscopique et la surface tangente aux ondelettes intermédiaires. Mais si ce principe est bien une hypothèse que l'on peut ou non accepter, et comportant d'ailleurs certaines incohérences, il reste qu'elle est étayée par un modèle à la fois physique et géométrique que ne possédaient pas les hypothèses de Descartes, Hobbes, Fermat ou Leibniz.
[76] HALL, « Two unpublished lectures of Robert Hooke », 1951, p. 221-222.
[77] BLAY, « Christiaan, Huygens et les phénomènes de la couleur », 1984.
[78] SHAPIRO, « Kinematic Optics », 1973, p. 245-258. Pour ce qui est de la propagation rectiligne, le modèle de Huygens prédit en effet que le corpuscule d'éther situé juste au bord d'un obstacle va émettre une ondelette hémisphérique vers l'avant qui va en partie se propager derrière l'obstacle. Mais que l'œil ne sera pas sensible à la propagation de ces ondelettes isolées puisque leur effet ne viendra pas être amplifié par celui d'ondelettes

de Hooke, c'est que – comme celles qui la précèdent – la théorie hugonienne laisse trop de phénomènes optiques inexpliqués, et même inabordés, pour que l'on puisse juger que sa démonstration efficace de la réfraction soit une preuve de la véracité de ses hypothèses relatives à la vitesse ou à la nature de la lumière. Tout au plus renforcera-t-elle leur plausibilité[79]. Aussi, malgré ce franc succès, l'hypothèse sur les vitesses posée par Huygens reste une hypothèse, certainement antérieure à l'élaboration complète de son système, d'ailleurs[80]. Et l'illusion qu'il produit sur nous d'une démonstration définitive de la décélération de la lumière vient certainement autant de l'élégante simplicité de son système, que du fait qu'elle est confirmée par la mesure que l'on connait aujourd'hui. Mais il faut toujours se garder de ces raccourcis rétrospectifs, qui nous font attribuer un surplus de bon sens à ces théories anciennes qui nous paraissent se rapprocher plus des nôtres que leurs concurrentes ; et bien se garder de condamner les autres sur un critère si trompeur. Car dans les cas de Descartes, Leibniz et Hooke, la plausibilité de l'hypothèse inverse d'une accélération de la lumière dans l'eau est tout autant renforcée par les succès à expliquer d'autres phénomènes optiques, que l'hypothèse de Huygens est renforcée par sa justification de la réfraction. Chaque théorie présentant de retentissants succès, tout en restant confrontée à de considérables nœuds de résistance, aura donc la valeur démonstrative que son lecteur voudra bien lui accorder, compte tenu de l'importance relative qu'il attribuera aux problèmes associés à ces succès et ces inconsistances.

Ainsi, peut-on encore imaginer que si une mesure de la vitesse de la lumière dans l'eau avait alors révélé son accroissement par rapport à celle dans l'air, la théorie vibratoire de Huygens sous la forme qu'on lui connait se serait effondrée ; mais qu'elle aurait pu se reconfigurer en une nouvelle théorie vibratoire, rejetant par exemple l'orthogonalité du front d'onde avec le rayon (comme le fait Hooke), ou introduisant tout autre propriété pour lors « omise » de ces ondulations[81], afin de concilier la nature ondulatoire avec une accélération de la lumière dans l'eau. Ce qui aurait encore entrainé l'impossibilité d'écarter rigoureusement l'éventualité d'une nature vibratoire de la lumière

Quoi qu'il en soit, ce n'est évidemment pas ce qui s'est produit : la comparaison des vitesses de la lumière dans l'air et dans l'eau a bien confirmé l'hypothèse de Huygens. Mais le fait qu'il ait fallu attendre plus de cent cinquante ans ce résultat avait alors permis à une théorie des projectiles de lumière de l'éclipser totalement. Qui plus est, cette théorie de la lumière, fondée sur des propositions de Newton, proposait une justification dynamique de la réfraction concluant à l'accélération de la lumière à son entrée dans l'eau. Nous montrerons pourtant que l'association de cette condition des vitesses à la nature ici attribuée à la lumière ne relevait toujours pas d'une nécessité.

## 4 Les théories des projectiles de Barrow et Newton.

### 4.1 Isaac Barrow

Car en effet, avant de formuler sa propre théorie de la lumière Newton s'est vu enseigner celle d'Isaac Barrow, qu'il a eu pour professeur à Cambridge[82]. Celle-ci, bien que difficile à cerner,

---

voisines. Cependant l'argument ne convaincra évidemment pas les partisans des théories des projectiles, qui verront dans l'absence de lumière derrière les obstacles une invalidation évidente de la théorie de Huygens.

[79] Conclusion à laquelle Huygens souscrirait d'ailleurs certainement lui-même. Puisqu'il se refusait à argumenter de la véracité de ses théories sur la base des phénomènes qu'elles étaient en mesure d'expliquer, mais se contentait de défendre prudemment que ces succès n'offraient à ses théories qu'un certain degré de « vraisemblance ».

[80] Dans le plan lapidaire que Huygens projette pour sa future *Dioptrique* en 1673 on voit par exemple que la première « difficulté » qu'il oppose à Descartes est que dans son système on ne voit pas « d'où viendrait l'accélération » de la lumière entrant dans l'eau (HUYGENS, « Projet du Contenu de la Dioptrique », 1916, p. 742).

[81] Périodicité, transversalité, polarisation, ou – plus radicalement encore – immatérialité.

[82] Newton succéda d'ailleurs à Barrow à la chaire lucasienne de mathématiques en 1669 ; année même où furent publiées les *Leçons d'Optique* de Barrow ; dont l'éditeur n'était autre que Newton.

pour ce qu'elle combine certains aspects des théories des milieux à d'autres associés aux théories de l'émission, considère toutefois explicitement la lumière comme étant composée de projectiles microscopiques se propageant à travers l'espace à une vitesse considérable[83].

Cette conception balistique de la nature de la lumière mise à part, la justification que Barrow proposait de la réfraction était identique presque en tout point à celle de Hobbes, qui l'avait très directement inspirée[84] : le « rayon lumineux » de Barrow est également un objet tridimensionnel – composé toutefois de corpuscules en véritable mouvement de translation depuis la source jusqu'au récepteur – limité par une ligne de front, marquant l'avancée de la propagation de la lumière et orthogonale par essence aux côtés du rayon lumineux épais. Ce rayon épais, rectiligne dans les milieux homogènes, peut cependant se courber au contact de surfaces réfringentes, du fait d'un changement progressif de sa vitesse. Car Barrow soutient en effet que la lumière se déplace plus lentement dans les milieux denses. Dès lors, le passage progressif du faisceau de projectiles depuis l'air vers l'eau, où leur vitesse est plus faible – et alors que la partie du faisceau se trouvant toujours dans l'air continue à se propager à grande vitesse – entraine selon Barrow une rotation sans déformation du front de propagation, s'accompagnant d'une courbure locale du faisceau lumineux en direction de la normale[85]. Et cette opinion quant au changement de vitesse de la lumière repose à nouveau sur l'analogie mécanique triviale à nos yeux – mais dont la présente étude souligne qu'elle ne l'était pas alors – que les milieux plus denses s'opposent plus que les milieux dilués à la propagation des projectiles de lumière. Juste avant la théorie optique de Newton préexistait donc au moins une théorie des projectiles lumineux dont celui-ci avait connaissance, et parfaitement à même de justifier cinématiquement de la réfraction, mais en s'appuyant sur l'hypothèse d'une plus grande vitesse de la lumière dans l'air que dans l'eau.

### 4.2 Isaac Newton

Et bien que les opinions qu'il exprime publiquement sur la nature de la lumière soient toujours nimbées d'ambiguïté, il est très manifeste que Newton favorise lui aussi l'hypothèse d'une lumière composée de projectiles. D'ailleurs, si la loi de la réfraction est présentée dans son *Opticks* comme un axiome – avant d'être finalement déduite d'une hypothèse plus fondamentale[86] –, c'est dans les *Principia* seulement qu'elle est complètement démontrée. De fait, dans une section dédiée à la dynamique des corps matériels minuscules – mais explicitement adressée aux opticiens – la trajectoire de corpuscules soumis à l'action d'une force uniforme, perpendiculaire à la surface séparant les milieux et sensible seulement entre cette surface et un plan parallèle et très proche d'elle, est présentée comme analogue à celle des rayons lumineux lors de la réfraction[87]. Soumis à l'action de cette force orthogonale au dioptre[88] et uniforme entre les plans $Aa$ et $Dd$, le corpuscule incident selon la direction $GH$ avec une vitesse $v_i$ suivra temporairement une trajectoire parabolique (imperceptible à l'œil), avant de ressortir de la zone avec une vitesse $v_r$ tangente à la parabole en $I$ (Fig. 7). On peut alors montrer, par simple application des lois de la dynamique newtonienne au cas d'une force constante et uniforme, que le rapport $\frac{\sin i}{\sin r} = \sqrt{1 + \frac{2.f.AD}{mv_i^2}}$ est nécessairement constant, puisqu'il dépend uniquement de la masse $m$ du corpuscule, de sa vitesse d'incidence $v_i$, de la valeur de la force $f$ et de l'épaisseur $AD$ de sa zone d'application[89].

---

[83] BARROW, *Lectiones XVIII*, 1678, p. 16. Il est à noter toutefois que ces corpuscules éjectés des corps lumineux peuvent parfois entrainer dans leur mouvement les corpuscules flottants du milieu dans lequel ils baignent. Néanmoins, il semble bien que ces corpuscules éjectés du corps lumineux poursuivront leur course à travers le milieu et que certains seront même à l'origine de la sensation lumineuse, au moins autant que les quelques corpuscules du milieu environnant qui auront aussi été mis en mouvement. Il semble surtout que cette interaction entre corpuscules des deux milieux permette à Barrow de justifier le retour vers la source d'un certain nombre de corpuscules lumineux, par rebond sur ceux du milieu environnant ; et de contourner ainsi l'objection classique faite aux théories des projectiles qu'une perte continue de corpuscules matériels par le Soleil devrait entrainer une diminution notable de sa masse.

[84] SHAPIRO, « Kinematic Optics », 1973, p. 179-188.

[85] BARROW, *Lectiones XVIII*, 1678, p. 14-16. Voir aussi le commentaire de SHAPIRO, *Ibid.*, 1973, p. 153-155.

[86] NEWTON, *Ibid.,* 1730, p. 5, puis p. 68-71. En l'occurrence l'hypothèse selon laquelle « les Corps réfractent la Lumière en agissant sur ses Rayons selon des Directions perpendiculaires à leurs Surfaces ». En effet, sans autre hypothèse que (1) celle-ci, selon laquelle les corpuscules lumineux sont soumis à une force uniforme et constante perpendiculaire à la surface, à laquelle ils ne sont sensibles que dans une épaisseur infime de matériau, et (2) celle plus implicite, selon laquelle les lois de la dynamique peuvent s'appliquer aux rayons de lumière, Newton propose une démonstration différente de celle des *Principia* (que nous évoquerons par la suite). Il part de l'expression supposée connue, découlant des principes de la mécanique newtonienne : $v_{\perp r} = \sqrt{v_{\perp i}^2 + v_{\perp 0}^2}$, où $v_{\perp i}$ est la composante de la vitesse perpendiculaire au dioptre (donc parallèle à la force accélératrice) au point d'incidence, $v_{\perp r}$ après traversée de l'épaisseur de matériau sur laquelle s'exerce la force, et $v_{\perp 0}$ après traversée de cette même épaisseur dans le cas particulier d'une incidence rasante. Rappelant ensuite que la composante $v_\parallel$ de la vitesse parallèle au dioptre (donc orthogonale à la force) restera constante tout au long du mouvement, Newton déduit –

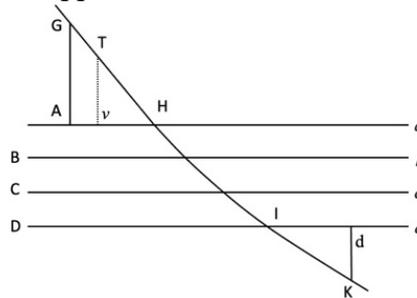

**Fig. 7 :** NEWTON, *Principia,* 1687, p. 229. Dans cet exemple, le rayon incident selon *GH* est dévié et ralenti en passant dans le second milieu, par une force orthogonale au dioptre *Aa* et n'agissant sur les projectiles de lumière que tant qu'ils sont dans la fine épaisseur *AD* du second matériau

Et ce n'est qu'une fois la constance du rapport des sinus démontrée, que Newton en déduit la constance et la valeur du rapport des vitesses[90] : puisque la force appliquée aux corpuscules est orthogonale au dioptre, la composante tangentielle de leur vitesse est conservée : $v_i.\sin i = v_r.\sin r$. Et le rapport des Sinus ayant par ailleurs été démontré constant, le rapport des vitesses l'est nécessairement aussi : $\frac{v_r}{v_i} = \frac{\sin i}{\sin r} = constante$. La constance du rapport des vitesses – qui jusque-là avait toujours été postulée, puis en quelque sorte confirmée par sa capacité à justifier la constance du rapport des sinus – devient une conséquence nécessaire de la constance du rapport des sinus ; elle-même démontrée indépendamment par l'étude de la déflexion par une force uniforme et constante d'un corpuscule soumis aux lois de la dynamique. Le modèle newtonien articule alors pour la première fois une explication simultanément géométrique, cinématique et dynamique de la réfraction. Explication si complète et rigoureuse

---

par une combinaison de rapports trigonométriques – que le rapport $\frac{\sin i}{\sin r}$ est nécessairement égal à une constante ; laquelle vaudrait en l'occurrence $\sqrt{1 + (\frac{v_{\perp 0}}{v_i})^2}$, même s'il ne l'explicite pas. Non plus qu'il n'explicite ici l'égalité de ce rapport avec celui des vitesses. Voir aussi SABRA, *Theories of light*, 1981, p. 305-308.

[87] NEWTON, *Principia*, 1687, p. 227-229. Ainsi que lors d'une réflexion, d'une réflexion totale, ou d'une inflexion (NEWTON, *Ibid.*, 1687, p. 231-235).

[88] La force orthogonale au dioptre est une nécessité tirée de l'observation (si tant est que l'on adhère au modèle dynamique) du simple fait qu'une force tangentielle produirait une réfraction même dans le cas d'une incidence normale ; et qu'une force de freinage, colinéaire à la vitesse, ne produirait pas de réfraction du tout.

[89] C'est en vérité une interprétation rétrospective de la démonstration newtonienne qui nous fait exprimer le rapport des sinus en fonction de $m$, $v_i$, $f$ et $AD$. On trouvera un commentaire rigoureux de cette démonstration et de sa version originale dans VILAIN, « Newton et le modèle mécaniste de la réfraction », 1987.

[90] NEWTON, *Ibid.*, 1687, p. 220-230. Traduction en français de la Marquise du Chastellet : « Les mêmes choses étant posées, la vitesse du corps avant l'incidence est à sa vitesse après l'émergence, comme le sinus d'émergence au sinus d'incidence » (Livre I, Proposition XCV, Théorème XLIX).

qu'elle ne laisse aucune place à l'interprétation et que l'on ne peut que s'y soumettre, si l'on souscrit par ailleurs aux principes mécaniques qui la sous-tendent[91]. Et cette démonstration conclue de manière rigoureuse et nécessaire à l'accélération de la lumière dans l'eau ; en accord avec Descartes, Leibniz et Hooke ; contre Hobbes, Fermat et Huygens. Ainsi, fort de sa conviction toute baconienne que les observations et expériences peuvent être réalisées indépendamment de toute hypothèse préalable – et permettre alors d'établir des théories vraies, libérées de tout élément hypothétique – Newton pourra affirmer en toute conviction avoir « démontré » que la loi des sinus est « précisément vraie », ainsi que toutes les conséquences qui l'accompagnent[92].

On sait bien aujourd'hui l'impossibilité de cette vision idéalisée d'une découverte scientifique – ou même d'une simple expérimentation – affranchie de toute hypothèse. Et l'on sait même combien ce programme est loin de se réaliser dans les propres théories mécanique ou optique de Newton, malgré la très grande robustesse de sa théorie et toutes les précautions qu'il emploie à n'énoncer et prouver les propriétés de la lumière que « par le raisonnement appuyé sur l'expérience[93] ». On réalise d'ailleurs d'autant mieux la beauté de l'exploit newtonien que l'on sait combien il assemble d'hypothèses aujourd'hui considérées comme fausses mais parfaitement plausibles alors. Et c'est justement là l'une des clés de la démonstration par Duhem de l'impossibilité de l'« *experimentum crucis* » en physique.

## 5 De l'impossibilité de l'« *experimentum crucis* » en physique selon Duhem

Pour étayer cette démonstration, Duhem invoque en effet deux arguments. Le premier[94] est que lorsqu'à partir de deux hypothèses physiques concurrentes il est possible de prédire deux conséquences bien distinctes et discernables par une expérience, la confirmation expérimentale de l'une des deux conséquences prédites, et donc l'infirmation de l'autre, ne saurait – contrairement au cas de la réduction à l'absurde en géométrie – conférer un statut de vérité à l'hypothèse dont la prédiction a été confirmée ; le physicien ne pouvant jamais être certain d'avoir épuisé l'ensemble des hypothèses imaginables et compatibles avec la prédiction, la validité de l'hypothèse confirmée reste suspendue jusqu'à la prochaine instance d'invalidation. Nous reviendrons sur cette question dans notre conclusion.

Le deuxième argument avancé par Duhem[95] est alors que l'expérience ne permettra pas même d'invalider définitivement l'hypothèse dont la prédiction a été démentie : la prédiction en question – comme l'expérience elle-même – repose nécessairement sur une théorie complète, composée d'un tel lacis d'hypothèses que l'expérience ne pourra jamais nous apprendre autre chose que l'invalidité du système complet de propositions dont l'articulation a mené à la prédiction et à la mise en place de son test. Ainsi, l'invalidation de ce qui est finalement une conséquence particulière d'une théorie complète, plutôt que d'une simple hypothèse, ne permettra jamais de déterminer laquelle précisément des hypothèses de cette théorie (si tant est qu'il n'y en ait qu'une) doit être remise en question – ni même si celle qui était initialement suspectée est nécessairement mise en cause. Dès lors, si la preuve expérimentale fournie par Foucault du fait que la vitesse de la lumière est inférieure dans l'eau à ce qu'elle est dans l'air invalide effectivement le *système* de l'optique newtonienne, elle n'invalide toutefois pas l'*hypothèse* des projectiles lumineux : « si les physiciens eussent attaché quelque prix à ce labeur, ils fussent sans doute parvenus à fonder sur cette supposition [celle des projectiles lumineux] un système optique qui s'accordât avec l'expérience de Foucault[96] ».

---

[91] C'est-à-dire si l'on accepte – au-delà de l'existence et de la forme de cette force réfractive définie par Newton – la possibilité d'une action à distance, la possibilité pour les projectiles lumineux de se propager à l'incroyable vitesse de la lumière, de se croiser sans se choquer, ou d'être émis en continu par une source comme le Soleil sans produire une perte notable de sa masse ; autant d'hypothèses qu'un pur philosophe mécaniste tel que Huygens par exemple ne saura accepter. Christiane Vilain fait d'ailleurs remarquer qu'en réalité la démonstration newtonienne de la constance du rapport des sinus fait intervenir le paramètre de la parabole parcourue par le corpuscule lumineux étudié et non directement la force qui lui est appliquée et la distance sur laquelle elle s'exerce. Qu'à ce titre, la démonstration est incomplète et ne peut rigoureusement constituer une preuve. Mais que sa formulation extrêmement persuasive suffit à rendre « réel et crédible » le modèle newtonien de la réfraction par attraction (VILAIN, « Newton et le modèle mécaniste de la réfraction », 1987, p. 323).
[92] NEWTON, *Opticks,* 1730, p. 68.
[93] NEWTON, *Ibid.,* 1730, p. 3. Voir par exemple la critique logique que Duhem fait d'une telle démarche, qu'il qualifie précisément de « méthode newtonienne » (DUHEM, *La théorie physique*, 1914, p. 289-304).
[94] DUHEM, *Ibid.*, 1914, p. 285-289.
[95] DUHEM, *Ibid.*, 1914, p. 278-285.

Dans cette veine, Alexander Wood[97] suggérait une modification de l'optique newtonienne qui aurait pu la réconcilier avec le résultat de cette expérience : selon lui, il aurait suffi de supposer que la lumière ralentirait à son entrée dans l'eau sous l'effet d'une force de friction tangentielle, alors que sa composante perpendiculaire serait conservée. Wood en concluait que l'expérience cruciale de Foucault n'avait alors pas véritablement invalidé la théorie des projectiles newtoniens, puisque celle-ci aurait pu s'adapter au résultat de l'expérience. Mais A. I. Sabra[98] a bien montré depuis l'inconsistance de cette contre-proposition de Wood, dont on est forcé de déduire la constance du rapport des cosinus et non des sinus, lors de la réfraction. Ainsi, Sabra – comme Duhem qui bien sûr ne s'y était pas trompé – confirme que l'expérience de Foucault a bien invalidé la théorie newtonienne de la réfraction, c'est-à-dire qu'elle a définitivement écarté la possibilité que les rayons de lumière marquent la trajectoire de projectiles matériels suivant les lois newtoniennes de la dynamique. Bien qu'elle ne soit effectivement pas en mesure d'écarter définitivement l' « hypothèse des projectiles » en général, ou l'ensemble des théories de la réfraction qui pourraient s'y rapporter : comme le suggèrent la théorie des projectiles de Barrow évoquée plus haut, dont les prédictions sont compatibles avec l'expérience de Foucault ; la possibilité au moins théorique d'associer un modèle de projectiles au principe du moindre temps de Fermat ; ou celle de justifier aujourd'hui de la loi de la réfraction par une théorie de la dynamique du photon[99], différente de la dynamique newtonienne.

## 6 Conclusions

Le panorama proposé jusqu'ici, qui s'achève à l'aube du XVIIIe siècle, semble d'abord indiquer que pour la plupart des auteurs évoqués, le rapport des vitesses de la lumière dans l'air et dans l'eau n'a pas été déduit logiquement d'une prémisse qui aurait été le choix de la nature de la lumière, à laquelle aurait ensuite été appliquée un modèle physico-mathématique complet de la réfraction, permettant de justifier rigoureusement le rapport constant des sinus. Mais que – à l'exception de Newton, peut-être – ces auteurs auraient dès le départ postulé conjointement (1) une nature particulière – mais toujours mécanique – de la lumière, et (2) un sens de variation de la vitesse de la lumière dans son passage de l'air à l'eau qui soit à leurs yeux mécaniquement compatible avec la nature postulée. Ce second postulat étant élaboré à partir du premier, sur la base d'une simple analogie mécanique avec le comportement supposé qu'aurait eu l'entité matérielle associée à la lumière (pression, projectile, fluide, vibration) lors de sa traversée de milieux plus ou moins « denses », « durs » ou « résistants ».

Seulement ensuite, les savants évoqués auraient-ils développé un modèle physico-mathématique rendant compte de la constance du rapport des sinus lors de la réfraction et compatible avec ces deux postulats. Et il semble bien que la liberté d'élaboration de ces modèles physico-mathématiques ait été suffisante pour que l'on ait toujours pu en développer un qui convienne à chacune des combinaisons envisageables des deux premiers postulats : que la lumière ait été envisagée comme une pression ou une vibration se propageant dans un milieu corpusculaire continu, qu'elle ait été considérée comme une entité empruntant les voies les plus aisées ou un flux de projectiles, il s'est trouvé dans tous les cas des auteurs considérant comme plus plausible son accélération au passage dans l'eau, et d'autres considérant que c'était le contraire. Et chaque option a pu être justifiée par un système physico-mathématique crédible.

---

[96] DUHEM, *Ibid.*, 1914, p. 284.
[97] WOOD, *In Pursuit of Truth*, 1927.
[98] SABRA, « A Note on a Suggested Modification of Newton's Corpuscular Theory of Light », 1954.
[99] DROSDOFF, WIDOM, « Snell's Law from an Elementary Particle Viewpoint », 2005.

Aussi, y a-t-il fort à parier que, si la vitesse de la lumière dans l'eau avait été mesurée un siècle ou deux plus tôt, chacune des hypothèses relatives à la nature de la lumière aurait pu produire au moins un système compatible avec le résultat de l'expérience. Et parmi celles invalidées par l'expérience, il en est même certainement qui auraient été suffisamment lâches et souples pour pouvoir se mettre en conformité avec ce résultat expérimental sans modification majeure de leurs fondements – en tout cas sans avoir à renoncer à une manière quelconque d'envisager la nature de la lumière.

Un siècle plus tard, il semble que les choses aient quelque peu changé ; qu'une décantation se soit produite dans les théories et les esprits, renvoyant dos-à-dos les hypothèses vibratoires et des projectiles relativement à la question de la vitesse de la lumière dans l'eau. Dans sa conférence Bakerienne *sur le Mécanisme de l'œil*, Thomas Young[100] défend prudemment l'idée que la lumière se propage plus vite dans l'air que dans l'eau. S'il se réfère pour cela à l'autorité d'une poignée d'auteurs passés tels qu'Isaac Barrow, Christian Huygens et Leonhard Euler[101], Young sait que cette opinion s'oppose à celle de « la doctrine la plus généralement admise » à l'époque, qu'est la théorie optique newtonienne[102]. Il n'est évidemment pas innocent que Young ait ici fait appel aussi bien à l'autorité d'Isaac Barrow – défenseur d'une théorie des projectiles, mais également maître de Newton – qu'à celles de Huygens et Euler – auteurs des théories vibratoires les plus avancées en son temps ; il fallait par tous les moyens et en tout point pouvoir défendre les conclusions de sa propre théorie vibratoire encore en développement. Mais ce qui se dessine néanmoins ici, c'est bien la bipolarisation des théories de la lumière qui en un siècle s'est développée autour de deux des grandes options mécanistes avancées quant à la nature de la lumière – vibrations du milieu et projectiles matériels –, et associée chacune à une conclusion opposée quant au rapport des vitesses de la lumière dans l'air et dans l'eau. C. Hakfoort associe cette bipolarisation à la pression exercée sur les partisans de l'optique newtonienne – qui au début du XVIIIe siècle semblait avoir balayé les théories concurrentes – par la publication de la

---

[100] YOUNG, « On the Mechanism of the Eye », 1801, p. 27.
[101] Nous avons évoqué la manière dont l'hypothèse d'une lumière plus lente dans l'eau s'accordait aussi bien de la théorie de l'émission de Barrow que de la théorie vibratoire de Huygens. Pour Leonhard Euler, les vibrations de l'éther se communiquaient plus ou moins aisément aux corpuscules de matière eux-mêmes ; et c'est cela qui différenciait un corps opaque ou réfléchissant d'un corps transparent. Les particules d'un corps transparent vibraient donc sous le coup des vibrations de l'éther, mais d'une vibration qui se propageait plus lentement du fait d'une densité et d'une élasticité différentes de celle de l'éther lui-même (EULER, *Conjectura Physica*, 1750, p. 11). Enfin, pour Thomas Young, cette idée que les particules des corps matériels elles-mêmes vibraient ne tenait pas. Ainsi défendait-il l'idée que l'éther pénétrait les corps matériels, et que du fait d'une attraction exercée par la matière sur l'éther, la densité de ce-dernier dans les corps matériels était d'autant plus importante que ce corps est dense : c'est alors cette augmentation de la densité d'éther qui selon lui justifiait mécaniquement d'une vitesse de la lumière plus faible dans les corps denses (YOUNG, « Outlines of Experiments and Inquiries », 1800, p. 128).
[102] Young renvoie non seulement à un passage de l'*Opticks* que nous avons déjà cité (NEWTON, *Opticks,* 1730, p. 5), mais aussi aux deux manuels d'optique les plus importants en Grande-Bretagne à l'époque que sont WOOD, *The elements of optics,* 1799 et SMITH, *A compleat system of opticks,* 1738 ; lesquels enseignaient exclusivement la doctrine newtonienne. Ce sentiment exprimé par Young d'une domination de la doctrine newtonienne au début du XIXe siècle est largement confirmé par les études de Cantor (*Optics after Newton*, 1983) centrée sur la Grande-Bretagne et l'Irlande, et de Chappert (*Histoire de l'optique ondulatoire,* 2007) centrée sur le cas français.

*Nova Theoria Lucis et Colorum* d'Euler[103]. C'est alors, d'après lui, que les théories dominantes de l'émission (rassemblées derrière le programme initié par Newton) et les théories renaissantes du milieu se retrouvent dos à dos ; que le débat en optique ne consiste plus qu'à devoir trancher entre ces deux options – tout argument contre l'une devenant un argument pour l'autre[104].

On comprend mieux comment, après réduction à l'extrême d'un panorama de théories optiques initialement bien plus vaste, Arago, puis Foucault, ont pu avoir le sentiment de pouvoir trancher entre l'hypothèse des projectiles et l'hypothèse vibratoire au moyen d'une seule expérience. Une circonstance est d'ailleurs venue se greffer à cela qui a pu favoriser l'illusion : si la théorie newtonienne avait définitivement fait le ménage dans le périmètre des théories de l'émission dès le début du XVIIIe siècle, plusieurs théories vibratoires avaient depuis réussi à développer un certain degré de crédibilité[105] ; mais bien qu'elles aient toutes été sensiblement différentes en un nombre de points relativement conséquents, celles-ci défendaient toutes exactement la même condition des vitesses – entretenant l'idée que l'option d'un ralentissement de la lumière dans l'eau était une conséquence nécessaire de quelque modèle vibratoire que ce soit, et l'autre une conséquence nécessaire de l'hypothèse des projectiles, qui ne reconnaissait de toute façon plus d'alternative au système de Newton. Il n'est donc pas surprenant que Foucault ait pu envisager qu'une comparaison expérimentale des vitesses de la lumière dans l'air et dans l'eau pourrait « servir à juger par une épreuve décisive les deux théories qui se disputent l'explication des phénomènes lumineux[106] » ; ni même que, dans un accès de militantisme, Arago ait affirmé pouvoir « décider, sans équivoque, si la lumière se compose de petites particules émanant des corps rayonnans […] ; ou bien si elle est simplement le résultat des ondulations d'un milieu[107] ».

Rétrospectivement seulement, la longue série de contre-exemples historiques proposée ici montre bien la vanité d'une telle entreprise : la versatilité des théories optiques, leur capacité à combiner modèles et hypothèses sans qu'aucune combinaison ne revête de nécessité absolue, confirmant l'impossibilité démontrée par Duhem d'une expérience cruciale susceptible d'invalider une hypothèse unique et isolée parmi la quantité d'hypothèses sur lesquelles repose une théorie physique. Tout autant qu'elle confirme l'incapacité totale pour une telle expérience – également annoncée par Duhem – de valider l'*hypothèse* ou le *système* ayant mené à la prédiction qui a été vérifiée : comme on a essayé de l'illustrer, les diverses conceptions scientifiques de la nature de la lumière défendues de 1637 à 1850 ont toutes trouvé des partisans de chacune des deux conclusions possibles de l'expérience de Foucault.

Ce que l'expérience de Foucault a pu invalider donc, ce sont des théories très localisées, prévoyant un résultat opposé à son résultat. Elle invalidait d'ailleurs ces théories d'autant plus efficacement qu'elles étaient étoffées, précises et structurées, donc qu'elles ne pouvaient aisément abandonner leur conclusion quant à la vitesse de la lumière dans l'eau sans devoir abandonner avec elle trop d'éléments qui y étaient associés ; à la manière du chêne de la fable de La Fontaine, les théories les plus élaborées ont perdu de la souplesse du roseau qui leur permettrait de plier selon le sens du vent, et elles se rompent toutes entières quand celui-ci se fait trop fort. Mais en même temps qu'une théorie s'est étoffée, elle a accumulé tant d'hypothèses interconnectées que la rupture du système ne permet plus de juger avec précision du point faible qui en a été la véritable cause. Ni même de savoir si sa faiblesse relevait d'une seule hypothèse ou de l'articulation de plusieurs. Non plus enfin que l'expérience aura pu exclure la possibilité que les théories qui n'ont pas été mises en défaut n'ont résisté au test que par la combinaison heureuse de plusieurs hypothèses erronées.

---

[103] EULER, *Nova Theoria Lucis et Colorum*, 1746.
[104] HAKFOORT, *Optics in the Age of Euler*, 1995.
[105] On pense évidemment à celles de Huygens, Euler, Young et Fresnel.
[106] FOUCAULT, *Sur les vitesses relatives de la lumière dans l'air et dans l'eau,* 1853, p. 16. C'est d'ailleurs à cette conclusion prudente que Foucault se contraint : « La conclusion dernière de ce travail consiste donc à déclarer le système de l'émission incompatible avec la réalité des faits » (FOUCAULT, *Ibid.,* 1853, p. 35).
[107] ARAGO, « Système d'Expériences », 1838, p. 954. Ainsi Arago offre-t-il ici un exemple paradigmatique de l'attitude dénoncée par Duhem. Ce dernier ne semblant rien à avoir à redire de la position de Foucault lui-même.

Dans le cas précis de l'expérience de Foucault on peut d'ailleurs constater que l'analyse de son résultat qui a pu être faite à l'époque reposait encore sur l'hypothèse mécaniste selon laquelle les comportements de la lumière ne pouvaient s'expliquer que par des modèles mécaniques inspirés de ceux élaborés pour la matière. Ce que permettait alors de conclure l'expérience de Foucault, c'est que dans le cadre strict d'une description mécaniste d'une lumière décrite en termes de corpuscules soumis aux lois des chocs ou du mouvement, les descriptions de la réfraction cartésienne (par la propagation instantanée d'une pression dans un espace plein), leibnizienne (par le choix de la voie la plus aisées), hookienne (par la vibration répétée de l'éther) et newtonienne (par la propagation de projectiles soumis aux lois de la dynamique) n'étaient plus tenables. Mais comme on l'a vu, comme l'histoire l'aura confirmé de manière étonnante au début du XX$^e$ siècle, et contrairement à l'affirmation originelle d'Arago, aucune conclusion définitive relative à la nature de la lumière ne pouvait être tirée de cette expérience.

Ce qu'elle permit toutefois indéniablement, c'est de faire basculer définitivement la charge de la preuve d'un camp à l'autre ; de reconfigurer à l'échelle d'une communauté les normes du plausible ; d'orienter en faveur d'un nouveau parti ce que Duhem appelle le *bon sens* – qui selon lui permet seul de décider entre deux options théoriques quand il est impossible de le faire selon les règles de la pure logique : « Après que l'expérience de Foucault eut montré que la lumière se propageait plus vite dans l'air que dans l'eau, Biot renonça à soutenir l'hypothèse de l'émission ; en toute rigueur, la pure logique ne l'eût point contraint à cet abandon, car l'expérience de Foucault n'était pas l'*experimentum crucis* qu'Arago y croyait reconnaître ; mais en résistant plus longtemps à l'Optique vibratoire, Biot aurait *manqué de bon sens*[108] ».

---

[108] DUHEM, *La théorie physique*, 1914, p. 331.

# Bibliographie citée